%% file: main.tex
\documentclass[sigconf]{acmart}

\AtBeginDocument{%
  }

\setcopyright{none}
\acmConference[Under review]{2025}{Global Perspectives of AI Risks and Harms}{2025}
\usepackage{wrapfig}
\usepackage{booktabs} % For professional table rules
\usepackage{geometry} % Optional: To adjust margins
\usepackage{tabularx}
\usepackage{colortbl}
\usepackage{colortbl}
\definecolor{codegreen}{rgb}{0,0.6,0} % You can leave this for code-specific colors
\definecolor{codegray}{rgb}{0.5,0.5,0.5} % You can leave this for code-specific colors
\definecolor{codepurple}{rgb}{0.58,0,0.82} % You can leave this for code-specific colors
\definecolor{backcolour}{rgb}{1,1,1} % Set background color to white
\definecolor{blackfont}{rgb}{0,0,0} % Ensure font is black

\usepackage{enumitem}
\usepackage{graphicx}
\usepackage{listings,newtxtt}
\usepackage{hyperref}
\begin{document}

\title{Global Perspectives of AI Risks and Harms: Analyzing the Negative Impacts of AI Technologies as Prioritized by News Media}

% RQ: What AI safety risks and harms does the media prioritize across countries and political bias?

\author{Mowafak Allaham}
\email{mowafakallaham2021@u.northwestern.edu}
\orcid{0000-0001-9211-4598}
\affiliation{%
 \institution{Northwestern University}
 \city{Evanston}
 \state{IL}
 \country{USA}}

\author{Kimon Kieslich}
\affiliation{%
  \institution{Institute for Information Law, University of Amsterdam}
  \city{Amsterdam}
  \country{The Netherlands}}
\email{k.kieslich@uva.nl}
\orcid{0000-0002-6305-2997}

\author{Nicholas Diakopoulos}
\orcid{0000-0001-5005-6123}
\affiliation{%
 \institution{Northwestern University}
 \city{Evanston}
 \state{IL}
 \country{USA}}
 \email{nad@northwestern.edu}

\renewcommand{\shortauthors}{Allaham et al.}
\renewcommand{\shorttitle}{Global Perspectives of AI Risks and Harms}
\begin{abstract}
  \input{abstract}
\end{abstract}




\received{22 January 2025}

\maketitle

\section{Introduction}
\input{introduction}
\section{Related work}
\input{related-work}

\section{Data}
\input{data}
\section{Methods}
\input{methods}
\section{Results}
\input{results}
\section{Discussion}
\input{discussion}\label{6}
\section{Conclusion}
\input{conclusion}
% \section{Ethics Statement}

% \begin{acks}
% \end{acks}

\bibliographystyle{ACM-Reference-Format}
\bibliography{references}
\newpage

\appendix
\section{Appendix}
\input{appendix}
\end{document}

%% file: abstract.tex
Emerging AI technologies have the potential to drive economic growth and innovation but can also pose significant risks to society. To mitigate these risks, governments, companies, and researchers have contributed regulatory frameworks, risk assessment approaches, and safety benchmarks, but these can lack nuance when considered in global deployment contexts. One way to understand these nuances is by looking at how the media reports on AI, as news media has a substantial influence on what negative impacts of AI are discussed in the public sphere and which impacts are deemed important. In this work, we analyze a broad and diverse sample of global news media spanning 27 countries across Asia, Africa, Europe, Middle East, North America, and Oceania to gain valuable insights into the risks and harms of AI technologies as reported and prioritized across media outlets in different countries. This approach reveals a skewed prioritization of \textit{Societal Risks} followed by \textit{Legal \& Rights-related Risks}, \textit{Content Safety Risks}, \textit{Cognitive Risks}, \textit{Existential Risks}, and \textit{Environmental Risks}, as reflected in the prevalence of these risk categories in the news coverage of different nations. Furthermore, it highlights how the distribution of such concerns varies based on the political bias of news outlets, underscoring the political nature of AI risk assessment processes and public opinion. By incorporating views from various regions and political orientations for assessing the risks and harms of AI, this work presents stakeholders, such as AI developers and policy makers, with insights into the AI risks categories prioritized in the public sphere. These insights may guide the development of more inclusive, safe, and responsible AI technologies that address the diverse concerns and needs across the world. 

%% file: introduction.tex
AI is becoming more integrated into various systems and applications that are serving millions of user globally. Despite its potential to drive innovation, economic growth, and increases in productivity \cite{mamasoliev2024impact,filippucci2024impact,technologyreview2023}, it poses significant risks to society, especially if misused \cite{bommasani2021opportunities,park2024ai,burtell2023artificial,de2023chatgpt,yang2023anatomy}. In response, governments, companies, and researchers have contributed regulatory frameworks \cite{madiega2021artificial,biden2023executive}, risk assessments \cite{solaiman_evaluating_2023, metcalf_algorithmic_2021,allaham_evaluating_2024,nanayakkara_unpacking_2021}, and safety benchmarks \cite{zeng_ai_2024,andriushchenko273323256agentharm,zhang2023safetybench} to govern \cite{reuel2024open}, anticipate \cite{kieslich_anticipating_2024, kieslich_my_2024, barnett_simulating_2024-1, bucinca_aha_2023, avin_exploring_2020, hautala_spectrum_2023}, and potentially mitigate such risks. However, identifying risks of emerging AI technologies remains difficult, for a variety of reasons including the complex entanglements and
interaction effects with the social worlds of human behavior and policy, particularly across different nations and societies, and especially during the pre-deployment phases of design and development \cite{bonaccorsi2020expert,herdel2024exploregen,allaham2024towards}. Without such information, AI developers cannot fully assess the potential safety concerns and risks of technologies they are developing for a global user base. In fact, prior research suggests that if technology developers had been aware of the impacts caused by similar technologies they are developing, many of these impacts could have been prevented through careful evaluations early in the design process of these technologies \cite{bruckman2020have,do2023s,pang2024blip}. 

In response, a few initiatives have emerged to organize, report, and monitor the observed risks and harms of AI technologies post-deployment in the form of incident databases \cite{mcgregor2021preventing,oecd_ai_incidents}. These databases document incidents based on the undesirable consequences of AI systems and technologies as articulated in news media (e.g, AI Incident Database, and OECD AI Incidents Monitor). While these databases are publicly accessible, they fall short in terms of (1) synthesizing global and region-specific risks associated with the deployment of AI technologies across different nations and societies, (2) articulating which risks are considered more critical by each nation, and (3) expressing how these risks may vary when considering the political biases surrounding risk perceptions \cite{paeth2024lessons}. Addressing these gaps contributes towards more inclusive risk assessments of AI that incorporate perceptions of these risks from regions that are usually under-represented, such as the Middle East and Africa, in the risk and safety evaluations of AI. Moreover, analyzing the interplay between the political bias of news outlets that are shaping the perceptions of these risks and the prioritization of these risks by different nations, may help mitigate future polarizing debates around AI risks, such as around economic harms \cite{huang2024uncovering}, that may influence the public perception of AI during the current and future development of AI regulations and policies. To address these issues, this research pursues the following research questions: \textbf{RQ1:} \textit{What are the categories of AI risks covered in global news media?} \textbf{RQ2}: \textit{What is the prevalence of these risks across global regions?} and \textbf{RQ3}:\textit{ How does the prevalence of these risks vary across regions when considering the political bias of the outlets reporting on them?}

Although news media reflects its own set of normative biases about what is prioritized for coverage and how AI is covered \cite{chuan2019framing,nguyen_news_2024}, it plays a crucial role in highlighting risks and harms relevant to the general public and shaping their perception of AI--a key stakeholder in shaping the current and future development of AI regulations and public policy in democratic countries. Accordingly, by analyzing a broad and diverse sample of global news media spanning 27 countries spread across six regions (Asia, Africa, Europe, Middle East, North America, and Oceania)%\footnote{Due to a limitation of our sample, this work does not include news coverage of AI by news outlets in Latin America}
, our research identified novel risks that are not as prominently covered by risk frameworks and safety benchmarks. In addition, it provides insights into the coverage of AI risks and harms reported across media outlets globally and at the regional level. Specifically, our findings reveal the global prevalence of 6 AI risks in our sample from news media coverage of AI: \textit{Societal Risks}, \textit{Legal \& Rights-related Risks}, \textit{Cognitive Risks}, \textit{Content Safety Risks}, \textit{Existential Risks}, and \textit{Environmental Risks}, reflecting how those risks are prioritized by media and thus help set the agenda for public perception. Furthermore, our findings highlight how the coverage of such risks varies based on the political bias of news outlets, which is crucial to account for to better shape more inclusive and globally informed AI regulations.

% Overall, this paper contributes:
By leveraging news media, our research provides some guidance to stakeholders and entities involved in the development of AI technologies within a country, but with potential plans to deploy them regionally or globally in a safe and responsible manner. In addition, this work offers a bottom-up approach that AI safety experts, and regulators can use to (1) understand and prioritize evaluations of certain risks and harms associated with AI technologies across different nations, (2) provide insights into potential partisan (mis)alignment with respect to various risks and harms of AI technologies which may be useful for politically-aware governance initiatives, and (3) expand existing impact assessment and AI safety frameworks to incorporate global and broader risks and harms associated with the use of AI technologies across different societies.

%% file: related-work.tex
Here we first critique current risk assessment practices, focusing on the strong emphasis on expert voices from the Global North, and a proposal of the analysis of news media coverage as a promising way to enrich the current assessment landscape. Thus, in subsection two, we discuss the role of media coverage in risk assessment. 

\subsection{AI Risk Assessments}
In order to safeguard the development and implementation of AI technology, risk assessment of AI has received increasing attention in recent years. Governments, companies, and researchers alike have published an enormous number of risk assessment reports and safety benchmarks (e.g., \cite{bird_typology_2023, weidinger2023sociotechnical, weidinger_taxonomy_nodate, dotan_evolving_2024, slattery2024ai, zeng_ai_2024}. While these frameworks uncover a variety of different risks, scholars have also critically interrogated current risk assessment practices \cite{kieslich_using_2024, hartmann_addressing_2024}. A major structural concern with current risk assessment practices is the allocation of power. Who defines risks in the first place and with what intentions. In particular, scholars have criticized first or second party risk assessments, i.e., technology developers and external parties hired by developing companies to conduct risk assessments or audits \cite{hartmann_addressing_2024, cohen_introduction_2023, floridi_ethical_2020, pasquale_power_2023}. Since the main goal of companies is to be economically successful, researchers fear that assessments by companies or their research teams tend to focus on those risks that are relatively easy to address and leave out other issues (e.g., socio-technical issues) \cite{griffin_what_2024}.

Related to this, scholars criticize a lack of inclusiveness in risk assessment practices. Often, risk assessment is based on the expert judgment of researchers or developers. However, while this expertise is certainly necessary to identify a plethora of risks, it is still limited to the professional view of these experts and fails to recognize risks outside their experience \cite{nanayakkara_unpacking_2021}. But some risks materialize only for users or affected groups and communities, especially for technologies that are highly shaped by user input, such as LLMs. Scholars have thus emphasized the importance of non-expert knowledge as an additional resource for risk detection \cite{metcalf_algorithmic_2021, moss_assembling_2021}. The lived experiences of a wider array of stakeholders can help to detect new harms that are not anticipated by experts, as well as to get a sense of the importance and materialization of these harms for affected people \cite{poortvliet_performativity_2016}. There is also a strong focus on countries in the Global North. However, people in the Global South have different challenges with AI, and consequently different risks may emerge that are not covered by current risk assessments \cite{roche_ethics_2022, wakunuma_africa_2024}. While scholars recognize the need to increase the inclusion of independent voices (including non-experts) \cite{metcalf_algorithmic_2021, moss_assembling_2021, krafft_how_2022, costanza-chock_who_2022}, few assessments actually do so (with some exceptions such as \cite{kieslich_anticipating_2024, kieslich_my_2024, barnett_crowdsourcing_2022,allaham2024towards}). In this study, we focus on a resource that is 1) structurally more independent from corporate practices, and 2) should (by quality standards, and when appropriately sampled) ensure a broader inclusion of societal perspectives: media coverage.

\subsection{The Role of Media Coverage in Risk Assessments} 
The media play an influential role in shaping the national and public discourse on AI, including by helping to set the standards and expectations for AI accountability \cite{diakopoulos_2025}. In the traditional understanding of communication science, the media function as agenda setters \cite{mccombs1972agenda}. A key task of the media is to inform a broad public about politically and socially relevant issues \cite{jamieson_how_2017} and thereby ensure a plurality of voices, i.e. the inclusion of all societal stakeholders. \emph{How} the media portray these technologies is critical, as media coverage has been shown to influence public opinion \cite{nisbet_knowledge_2002, scheufele_public_2005}, especially for novel technologies like AI with which ordinary citizens have little personal experience \cite{jamieson_recap_2017}. Public opinion thus plays a crucial role in technology adoption. On the one hand, citizens act as consumers of AI technology, and the media's portrayal of AI can influence whether or not people are willing to use the technology \cite{ouchchy_ai_2020}. On the other hand, citizens can act as voters and thus influence regulatory aspects \cite{ouchchy_ai_2020, kieslich_ever_2023, kieslich_role_2024}.

Recognizing the importance of the news media in relation to AI, a significant number of scholars have focused on analyzing how the news discusses AI. Most studies of media coverage focus on media coverage from countries in the Global North, such as the US \cite{chuan2019framing, fast_long-term_2017}, the UK \cite{brennen_industry-led_2018, brennen_what_2022, roe_what_2023}, Germany \cite{meisner_risks_2024, kieslich_everything_2022}, the Netherlands \cite{vergeer_artificial_2020}, or take a comparative approach between the US and the UK \cite{nguyen_news_2024, bunz_ai_2022} or the US and China \cite{nguyen_new_2022}. Only a few studies focus on non-Western countries, with the exception of China \cite{zeng_ai_2024} and a comparative study of 12 countries, including countries of the Global North and the Global South \cite{ittefaq_global_2025}. Thematically, studies on media coverage mostly focus on mapping the general discourse on AI, for example by analyzing the thematic structure or sentiment of media discourse (e.g. \cite{ittefaq_global_2025, brennen_industry-led_2018}), while rarely focusing explicitly on risks or negative impacts (with the exception of \cite{ouchchy_ai_2020, nguyen_news_2024,chuan2019framing}).

One of the main principles of journalistic news quality is to represent a plurality of voices that are relevant to the discourse, and the inclusion of these voices (e.g., activists, academics, civilians, NGOs) can enrich the discourse on AI \cite{brennen_industry-led_2018}. In particular, when it comes to reporting on AI risks, the efforts of investigative journalists have helped shed light on pressing issues such as the child benefit scandal in the Netherlands \cite{constantaras_inside_2023} or the COMPAS recidivism algorithm in the US \cite{angwin_machine_2022}. Indeed, scholars first articulated the idea of ``algorithmic accountability'' as stemming from investigations published in the news media \cite{diakopoulos_2015}, and more recently have argued for the inclusion of journalists in third-party audits of AI systems, as they ``were responsible for uncovering deeply-rooted socio-technical harms in algorithmic systems related mainly to representational harms due to discriminatory design choices.'' \cite{hartmann_addressing_2024}. % As a result, they showed the extent of the potential harms brought by AI systems’ biases
This is supported by the fact that many of the sources of the AI Incident Database \cite{mcgregor2021preventing} are newspaper articles. As a result, news coverage of AI risks plays a key role in exposing the risks of AI systems. Unlike self-reporting by companies (including their research teams), journalists are structurally independent and can uncover novel impacts that may conflict with corporate goals. Their inclusion ``ensure[s] social accountability through domain knowledge and special access to affected communities'' \cite{hartmann_addressing_2024}. %Additionally, journalists have demonstrated their importance in financial and algorithmic audits as an accountability component in the past. 

Another important factor when considering the impact of discourse on public perception is its politicization. Scholarly research in this area states that politicization of an issue requires three conditions \cite{de2011no, schattschneider1957intensity}: (1) polarization of the issue, i.e., whether and how prevalent different (political) positions are on the issue. This could also be achieved by different framing or agenda setting of topics related to the issue (e.g. different prevalence of AI risks). (2) The intensity of media coverage. This refers to the visibility of an issue. The more it is covered, the more relevance is attributed to the issue. And (3) The resonance of the issue, i.e. how relevant the issue is in the eyes of the public. Media coverage plays a key role in this regard, as it provides an important arena in which AI is discussed. Several studies of media coverage have found a sharp increase in news coverage of AI in recent years \cite{fast_long-term_2017, vergeer_artificial_2020, ouchchy_ai_2020, ittefaq_global_2025, chuan2019framing}, which satisfies the condition of intensity of coverage. Several scholars have also analyzed the influence of the political leaning of the news outlet on the framing of AI -- with mixed results \cite{brennen_industry-led_2018, roe_what_2023, vergeer_artificial_2020}. However, the politicization of the AI \emph{risk debate} in particular has not been explored. Analyzing the politicization of AI risks in terms of political positions is important because it reflects political strategies in terms of regulation or policy enforcement. Furthermore, it shows how citizens who consume politically biased news are informed and perceive the AI risk discourse.

Overall, we find that there is a gap in the use of news sources in risk assessment practices and that  literature on news analysis doesn't engage with AI risks. Therefore, in this paper, we address these gaps by (1) using news as a source to inform risk assessment, (2) explicitly focusing on AI risks in the study of news content, (3) taking a global perspective in analyzing news from countries in the Global North and Global South, thus contributing to a more diverse analysis of AI risks, and (4) analyzing the effect of political positioning of news coverage on the prevalence of AI related risks.

%% file: data.tex
\textbf{Curating AI-related articles from news media}. To establish a dataset of online news articles related to AI from around the world, we used GDELT \cite{leetaru2013gdelt} (Global Data on Events, Location and Tone Project) to collect online news articles published in English between January 2022 and October 2024 based on a previously published list of 41 AI-relevant keywords in English (listed in Appendix \ref{a3}) sourced from news media \cite{allaham2024towards}. GDELT captures extensive coverage of what is reported on in news media across different news outlets, languages, and offers an accessible way of querying such coverage via an API \cite{leetaru2013gdelt, ward2013comparing}. In addition, it presents an alternative to scraping news portals and aggregators, such as Google News, that display or rank news content that is recently published by popular news outlets, politically biased (i.e., slight leftward bias), or limited in exposure to global perspective \cite{nechushtai2024more,ulken2005question,hernandes2024auditing}.
To retrieve AI-related articles from GDELT, we constructed multiple API requests in Python using GDELT's v2 API endpoint\footnote{\url{https://api.gdeltproject.org/api/v2/doc/}}. For each of the 41 AI-related keywords, an API request was sent to GDELT to retrieve the URLs and metadata of the daily published news articles mentioning that keyword. In total, we retrieved all URLs corresponding to 1,218,058 online articles published between January 2022 and October 2024 by 21,383 unique news outlets from 30 countries around the world.

\textbf{Filtering for national news domains}.
To analyze the risks of AI prioritized in news coverage across different countries, we refined our data by filtering URLs to include only those from national news domains in each country. This step is necessary to ensure that our sample is free from content published by corporate or technical blogs. To this end, we leveraged the Global English Language Sources \cite{mediacloud2025} from Media Cloud (MC) \cite{roberts2021media}. MC's list of 1,064 national news outlets that publish articles in English across 177 countries was used to filter the domains of articles we collected from the previous step. After applying the filter, we are left with 235,636 URLs with 600 domains from 30 countries.

\textbf{Domain bias rating}\label{3.3}
Although the global representation of news media in our sample is useful for understanding patterns of media coverage of negative impacts of AI, prior research has shown that the political bias of news domains tend to influence the discourse around scientific topics such as climate change \cite{chinn2020politicization,anonymous2024} or emerging technologies (e.g., nuclear energy), including AI \cite{brennen_industry-led_2018, roe_what_2023, vergeer_artificial_2020}. To rate the domains in our sample, we used Media Bias Fact Check (MBFC) domain-level ratings \cite{mediabias2025}. MBFC is an independent website maintained by researchers and journalists that relies on human fact-checkers affiliated with the International Fact-Checking Network to evaluate media sources along different dimensions such as factual reporting and bias \cite{lin2023high}. MBFC categorizes news sources in one of nine bias categories: least biased, left bias, left-center bias, right-center bias, right bias, conspiracy-pseudoscience, questionable sources, pro-science, and satire \cite{mediabias2025}. After aligning domains in our sample with MBFC ratings and excluding those categorized as questionable, conspiracy-pseudoscience, pro-science, or satire--as these domains focus more on factuality and type of content than political bias--our refined sample includes 277 domains, constituting 163,314 URLs to articles ($\sim$69.3\% of the filtered articles per the previous step).

\textbf{Scraping news articles}\label{3.4}. For each URL, we attempted to scrape the article text and title using a custom web scraper in Python that leverages the \texttt{newspaper} library \cite{newspaper2025}. We could not scrape a majority of the articles  due to pay- or sign up walls or missing content (i.e., 404 errors). As a result, the final sample containing the scraped articles includes 42,853 ($\sim$26.2\% of 163,314) articles published by 168 domains spanning 27 countries (see distribution of articles per country and global region in Table \ref{tab:tab3}).

%% file: methods.tex
\subsection{Summarizing negative impacts of AI from news media using an LLM}\label{4.1}

\textit{\textbf{Filtering articles by content}}. For the filtering step, we follow a similar zero-shot prompting approach to the one reported in previous research with a similar objective of detecting the negative consequences of AI in news media \cite{pang2024blip}. To develop our prompt, we referred to prior research on social impact assessments \cite{becker_social_2001,nanayakkara_unpacking_2021} to synthesize the following conceptual definition of an impact of an AI technology that we used to steer the LLM: \textit{An impact refers to an effect, consequence, or outcome of an AI system (i.e., model or application) that positively or negatively affects individuals, organizations, communities, or society}. We did not limit the conceptual definition to negative impacts per se, so as to provide opportunity for future work that may want to focus on positive impacts of AI technologies \cite{kieslich_my_2024}.

To assess the ability of the LLM for classifying articles reporting impacts of AI, we randomly sampled 300 articles from our corpus. One of the authors annotated each article to determine whether it contained at least one impact of AI, based on our definition and found that most articles (77\%) had an impact. Using this annotated sample and prompt \ref{p1}, we used GPT-4o to classify each article as either containing an impact or not. The model performed well on this task achieving an F1-macro score of 0.82 (macro-averaged precision=0.81 and macro-averaged recall=0.83). We then applied this prompt using the OpenAI batch API to the rest of the dataset. Out of the 42,853 articles in our sample, as reported in section \ref{3.4}, GPT-4o classified 32,439 (75.69\%) as containing impacts.

\textit{\textbf{Summarizing negative impacts}} After identifying 32,439 articles as having impacts of AI, we instructed GPT-4o with prompt \ref{p2}, including the entire article text as context to the model, to summarize all the negative impacts reported in each article. Each article could contain multiple impacts, and each was represented separately in a list. This process resulted in 47,731 negative impacts of AI that were summarized from 16,312 articles sourced from 157 domains spanning 27 countries.

\subsection{Annotating negative impacts from news media}\label{4.2}
\textit{\textbf{Human annotations of negative impacts}}.
Due to the large number of negative impacts summarized from articles as outlined in the previous section, we explored clustering these impacts and annotating the resulting clusters by sampling and annotating impacts from each cluster (as described in details in Appendix \ref{clustering-method}). However, we observed that the impact statements within each cluster refer to different types of impacts. Accordingly, we modified the annotation process so each summary of impact statement is annotated based on the type of impact, rather than the contextual use of AI per the outcome of the clustering approach, as an AI technology could have more than one negative impact within the same contextual use which could provide us with the range of impacts necessary for subsequent analysis. To do so, we 1) involved two authors to independently annotate a randomly selected sample of negative impacts, and then 2) calculate the inter-coder reliability to confirm the validity of the annotated impact categories.

Based on the AI Risk Categorization taxonomy (AIR-taxonomy) \cite{zeng_ai_2024}, two authors independently annotated a total of 1,060 negative impacts. Although other expert-driven taxonomies of AI risks exist \cite{slattery2024ai,solaiman_evaluating_2023,shelby2023sociotechnical,weidinger_taxonomy_nodate}, prior research found that these taxonomies may suffer from inadvertent expert or selection biases and may not be as representative of global perspectives of AI risks and harms \cite{allaham2024towards, bonaccorsi2020expert, crawford2016artificial, hagerty2019global,jobin_global_2019}. Alternatively, the AIR-taxonomy presents a global, yet limited, perspective into the risks and harms of AI that is derived from eight government policies, including the European Union, United States, and China, and 16 company policies worldwide. This global representation of risks, at least by governments and industry, is more aligned with our research objective of incorporating global perspectives of AI risks and harms into the impact assessment process of AI. 

For each impact summary in the sampled impacts, annotators categorized the impact summary into one of the 16 Level-2 categories of risks presented in the AIR-taxonomy \cite{zeng_ai_2024}. We decided to code summaries of impacts at Level-2, because it offers a balanced level of granularity of the risk categories that are grouped based on societal impact \cite{zeng_ai_2024} and enable us to re-organize the annotated impacts into Level-1 for subsequent analyses and comparisons between regions. If an impact summary aligns with one of the AIR-taxonomy's Level-2 subcategories of risks, it is annotated with that category's label. Alternatively, an impact summary is assigned an ``other'' label. Next, per the constant comparison process in qualitative thematic analysis \cite{braun2012thematic}, each impact summary in the ``other'' category was assigned to an emerging category of risks. The definitions corresponding to the emerging categories were continuously revised and assessed throughout the coding process to ensure that these categories made conceptual sense, and then decide whether they needed to be reorganized, merged, or further broken into more categories as suggested by qualitative research methods \cite{braun2012thematic}. After coding all 1,060 impact summaries, the categories were largely saturated and stable showing no new emerging categories of risks.

To evaluate the reliability of the annotated impacts by human coders based on the AIR-taxonomy, as well as on the emerging categories, we calculated Kripendorph's alpha \cite{marzi2024k}. The metric showed a high inter-coder reliability (K-alpha=0.97) between the two annotators, indicating that the definitions of the impact categories are consistent and stable. The distribution and prevalence of each impact category in the sample can be found in Table \ref{tab:tab1} in the Appendix. The list of the risk categories and their definitions is provided in Appendix \ref{definitions-of-impacts}, which is also utilized to annotate the remaining negative impacts in our corpus using an LLM.

\textit{\textbf{LLM annotations of negative impacts}}.
To scale up the annotation process using an LLM, we had to first validate the efficacy of the LLM in annotating impacts to their corresponding categories based on the sample of 1,060 annotated impacts. Accordingly, using a zero-shot prompting approach, we sent requests via the OpenAI API instructing GPT-4o to annotate each impact summary for only one of the defined categories by the annotators, as described in the previous section and outlined in Prompt \ref{p3}. In addition, we included in the prompt an instruction for the LLM to assign an ``other'' label for impacts that do not fit any of the categories defined in the prompt, mirroring the best practices of thematic analysis \cite{braun2012thematic}. We also included a ``no\_impact'' category to capture potential false positives in the data--summaries from articles that do not fit our definition of a negative impact.

Using the human annotated categories of the 1,060 impacts as a baseline, we evaluated the efficacy of GPT-4o on this task after aligning the annotated categories back to Level-1 of AIR-taxonomy, focusing our analysis at the category rather than sub-category level, to facilitate a more interpretable comparison of risk categories across regions. While the model achieved a suboptimal macro averaged F1-score of 0.60, it is comparable to results observed in other multi-class classification tasks \cite{pang2024blip,rao2024quallm}. Despite that, and based on the classification report presented in Table \ref{tab:tab2}, we observed a few categories having a decent recall score of 0.70 or higher but are also associated with low precision scores (i.e., low to mid sixties). We focused our evaluation on recall scores, rather than F1-scores, to assess the LLM's ability in capturing the majority of relevant instances within each category. This decision was influenced by the LLM's poor performance on certain categories, such as Structure/Power, which impacts the overall F1-score. By manually reviewing instances of categories that had a recall score of 0.7 or higher, such as Cognitive Risks or Content Safety Risks (as shown in Table \ref{tab:tab2}), we observe that when the LLM misclassifies an impact, it predominately classifies it into other plausible impact categories, from the 31 categories of risks articulated in the previous section, rather than the category assigned by the two annotators. This has also been observed in a similar tasks focusing on extracting concerns regarding AI and algorithmic platform decisions from discussion forums \cite{rao2024quallm}. Furthermore, our manual review also shows around 14.6\% (155 out of 1,060) of impact summaries were misclassified by GPT-4o into a wrong and irrelevant category, affirming that the impacts with the recall scores of 0.70 or higher but low precision (i.e., low to mid sixties) can still contribute to the analysis.

Based on the classification report presented in Table \ref{tab:tab2}, and the sensitivity score of the LLM on each category, we selected six impact categories that had at least a recall score of 0.7 or higher for further analysis: \textit{Societal Risks, Legal \& Rights-related Risks, Cognitive Risks, Content Safety Risks, Existential Risks, Environmental Risks}. The prevalence of these categories of risks across geographical regions and political ideologies is analyzed in the results section.

%% file: results.tex
\begin{figure*}
\centering
\includegraphics[width=0.75\textwidth]{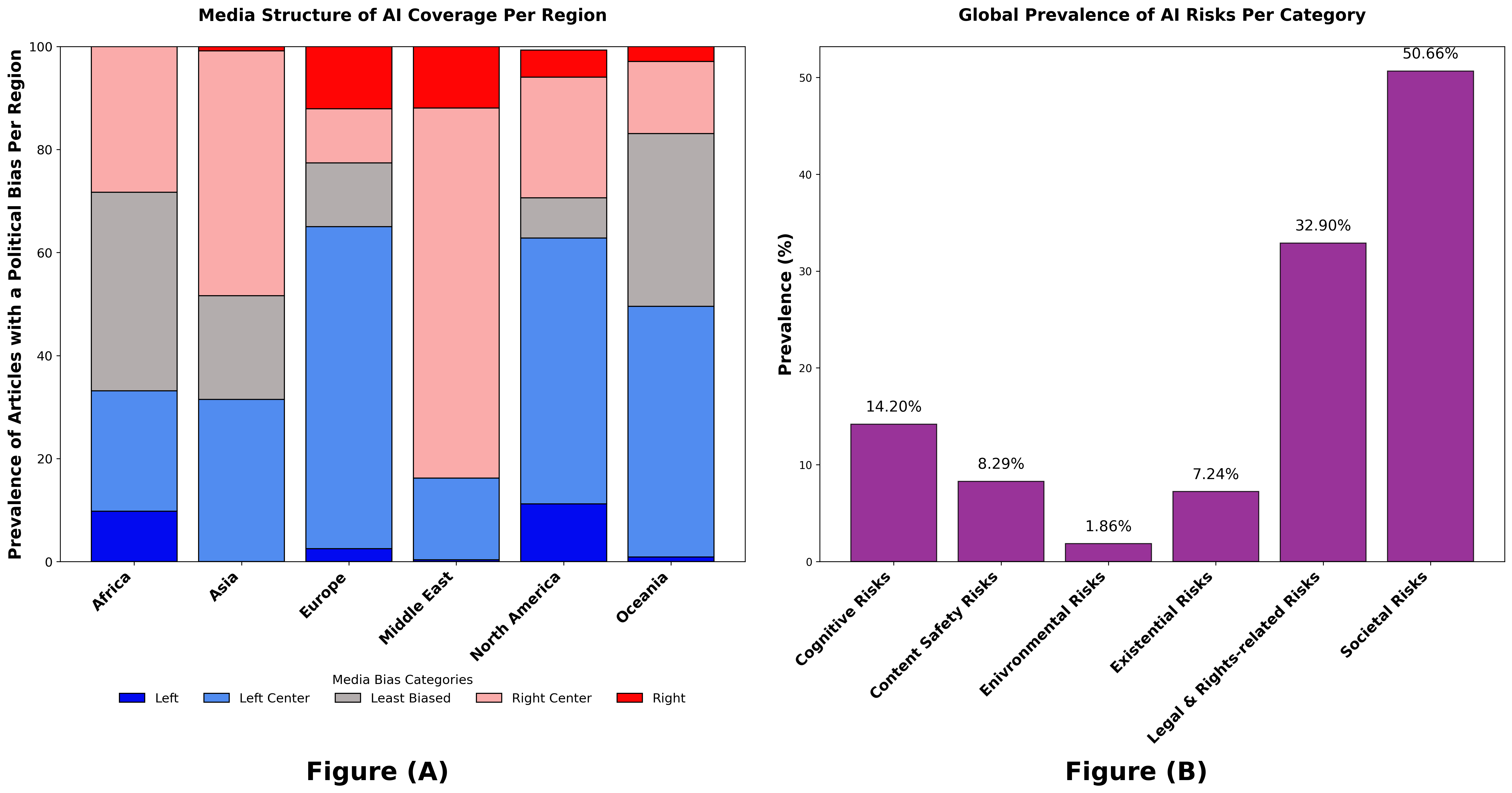} % Path to the image file
\caption{Figure (A) provides an overview of the media structure of AI coverage across regions stratified by political bias. News coverage of AI in our sample is predominantly driven by news outlets with left center bias. Figure (B) shows the global prevalence of AI Risks for each category. To calculate the global prevalence, we counted the number of articles reporting risks related to each category and divided each number by the total number of articles across regions and categories (i.e., 16,312).}
\label{fig:media-structure}
\end{figure*}

\subsection{Emerging AI risk categories from news media}\label{5.1}

The global representation of news media coverage of AI in our sample captures a broad range of risks and concerns about AI across different nations, enabling the expansion of the AIR-taxonomy with the addition of new categories. By annotating these risks and concerns in a sample, as described in Section \ref{4.2}, we contribute 16 risk categories, marked with asterisk in Table \ref{tab:tab1} and defined in \ref{definitions-of-impacts}, that are not present at any level of the AIR- taxonomy \cite{zeng_ai_2024}. These 16 risk categories, which emerged from annotating our sample, include: \textit{AI Governance Risks, Authoritarian Use of AI Risks, Disruption of Service Risks, Environmental Risks, Ethical Risks, Existential Risks, Humanness Risks, Information Risks, Media Risks, Mental \& Emotional Risks, Over-reliance Risks, Performance Risks, Safety Risks, Structure \& Power Risks, Technology Adoption Risks, and User Experience Risks}. All emerging categories were aligned back at Level-1 of the AIR-taxonomy to enable an analysis at the category rather than sub-category level, which facilitates a more interpretable comparison of risk categories across regions.

To align these emerging risk categories with the AIR-taxonomy of AI risks, one possible way is to organize the categories based on their contextual independence and their potential in addressing gaps in the taxonomy through elaborating on existing categories. For instance, \textit{Authoritarian Use of AI} and \textit{Media Risks} (i.e., risks of AI on the news and media industries) can possibly be integrated as Level-2 categories under \textit{Societal Risks}. \textit{Performance Risk}, \textit{Safety Risks}, and \textit{Disruption of Service Risks} could be grouped under the existing category of \textit{System and Operational Risks} at Level-1. In contrast, \textit{Mental \& Emotional Health}, \textit{Humanness Risks}, and \textit{Over-reliance Risks} did not fit any of the existing categories in the AIR-taxonomy and all share cognitive implications to using or interacting with AI, therefore one possibility is to aggregate them under a new category at Level-1 named \textit{Cognitive Risks}. In addition, \textit{Usability Risks} could encompass the two emerging categories related to \textit{Technology Adoption Risks} and \textit{User Experience Risks}. All remaining emerging risk categories-- \textit{Information Risks, Ethical Risks, Existential Risks, AI Governance Risks, and Structure/Power Risks}--were left as independent risk categories at Level-1, as we did not find them fit in any of the existing categories of the AIR-taxonomy. While our grouping of these categories introduce an expert bias to the analysis, due to our interpretation of these risks in our sample, we acknowledge that these risks can be grouped in various ways. Thus, we encourage future research in adopting a more theory-driven approach to grouping these categories by drawing from the impact assessment literature to understand their fit across a wider array of typologies. 

As a result of organizing and aligning these emergent categories of risks, we suggest 8 potential AI risks categories to the AIR-taxonomy at Level-1, bringing the total number of risk categories to 12, as listed in Table \ref{tab:tab4} in the Appendix. However, as outlined in the annotation step in Section \ref {4.2}, and based on the classification report in Table \ref{tab:tab2}, we choose to only include 6 of the 12 risk categories in subsequent analyses due to the model's satisfactory classification performance on these categories: \textit{Societal Risks, Legal and Rights-Related Risks, Cognitive Risks, Content Safety Risks, Existential Risks, and Environmental Risks}. Next, we elaborate on these six main risks categories.

\textit{Societal Risks} - features risks of AI that have negative societal implications on the public through AI-generated deception or manipulation. Other risks include the implications of AI on politics such as influencing elections by ``producing election misinformation about ballot deadlines'' or ``disseminate[ing] fake news". Other societal risks of AI have implications on the economy, which involves dis-empowering workers by potentially ``replace[ing] or take[ing] over jobs'' or may cause instability to financial systems if AI starts to develop ``complex financial mechanisms that may become too sophisticated for any human to understand, potentially leading to a loss of control over the financial system'' or ``distorting market signals in high-frequency trading''. Lastly, AI also has implications on news media such as ``weakening journalism'' and ``undermining the [public] confidence in the media'' due to the ease of ``creating fake content with AI''.

\textit{Legal and Rights-Related Risks} - reflect risks of AI resulting from privacy violations, due to ``AI voice cloning technology exploiting personal data from social media to create voice replicas'', or the embedded bias in AI causing LLMs to ``sexualize women in AI-generated images''. Furthermore, other risks in this category raise concerns related to AI committing copyright infringement ``to articles from news media'', or engaging in criminal activities to influence ``legal proceedings through tampering with evidence using deepfakes''.

\textit{Cognitive Risks} -- in this category, risks covered some cognitive implications of interacting or using AI systems, such as mental and emotional risks resulting from AI generating images that ``distort body images presenting an unrealistic aesthetic ideal for both men and women'', negatively impacting ``one's self-worth''. In addition, this category includes risks pertaining to the over-reliance on AI like the potential ``reliance on AI for decision making'' may result in ``over-reliance and trust in AI over human judgment''. Another key risk is the potential loss of human creativity and authenticity, as covered  by the humanness risks, due to AI ``undermining the unique human elements of creativity and imagination''. Furthermore, there were concerns of losing human touch because of ``[AI-generated] content that lacks the human element and nuances of real human relationships, leading to less fulfilling interactions''.

\textit{Content Safety Risks} - risks in this category are primarily associated with the potential of AI for generating ``child abuse material'' that causes child harm, or toxic speech such as generating ``sexually harassing messages'', or even explicit and non-consensual sexual content similar to the ``AI-generated nudity of Taylor Swift''. Also, this category includes risks related to the potential dissemination of sensitive content by LLMs related to ``building chemical or biological weapons", or violent and extreme content similar to Sarai, the chat bot that encouraged Jaswant Singh Chail to ``kill the queen of England''.

\textit{Existential Risks} - risks in this category predominantly express the potential existential threats and harms posed by AI on humanity such as ``demising of the human race''. Other risks include the potential loss of control over AI systems resulting in ``war, cyber conflicts, and deployment of nuclear weapons''. Moreover, risks of AI misalignment were also covered in this category due to concerns of AI ``counter[ing] human interests'' posing ``a significant existential threat'' to humanity.

\textit{Environmental Risks} - encompass environmental risks resulting from the considerable energy and water resources for the infrastructure and compute required to train LLMs and manufacture microchips (e.g., GPUs). This includes ``a rise in carbon emissions'' as AI use is ``increased across all sectors'', environmental and ecological degradation such the loss of biodiversity as a result of ``extracting rare earth metals'' for manufacturing AI hardware components ``lead[ing] to environmental degradation and biodiversity loss''.

\subsection{Global \& Regional Prevalence of AI Risk in News Media}

%%%%%%%%%%%%%%%%%%%%%%%%
\begin{figure*}
\centering
\includegraphics[width=0.9\textwidth]{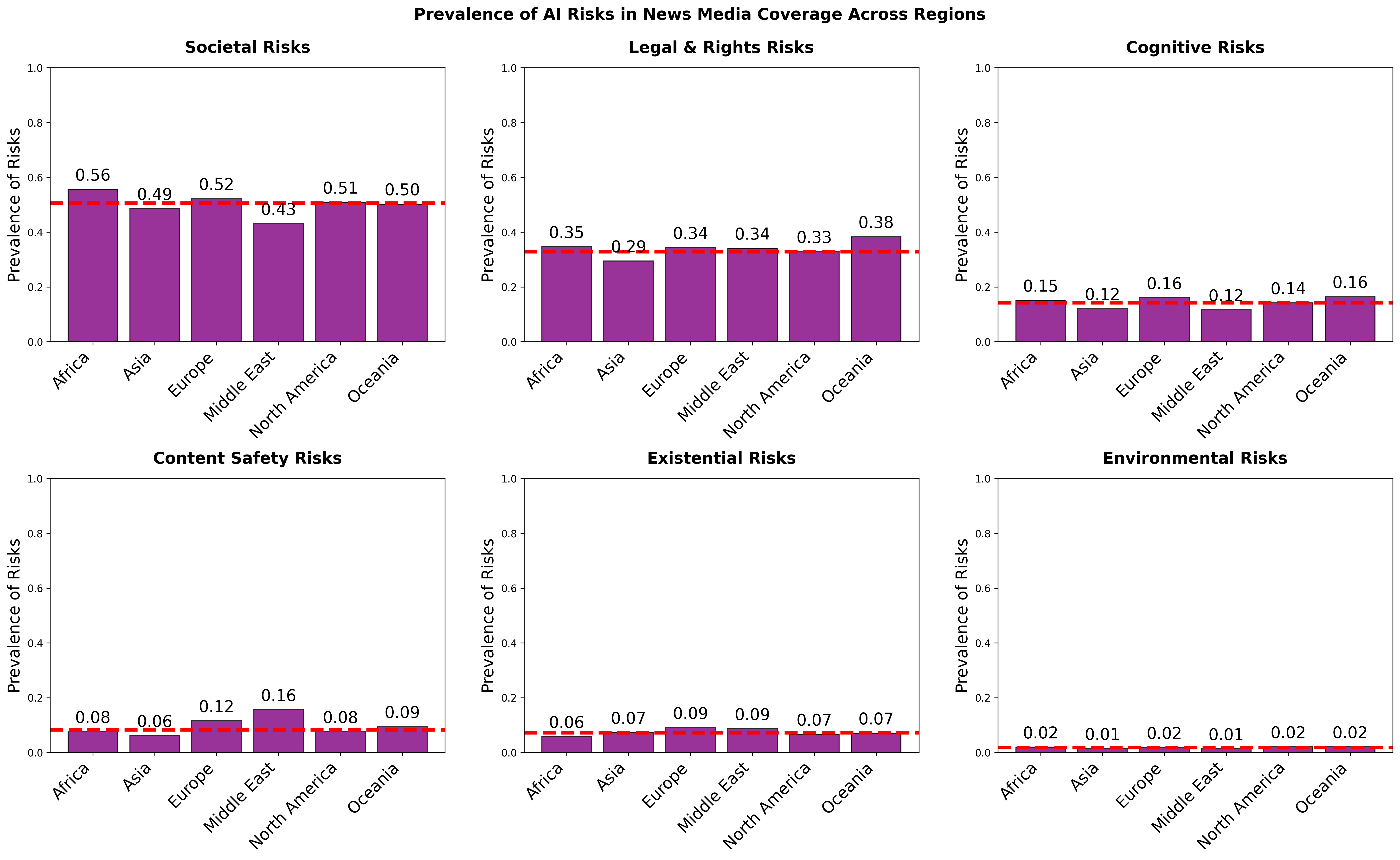} % Path to the image file
\caption{Prevalence of AI risks in our sample from news media coverage across six different regions. The dashed red line represents the global average prevalence per risk category.}
\label{fig:prevalence-pivto}
\end{figure*}
%%%%%%%%%%%%%%%%%%%%%%%%%

\subsubsection{Global Prevalence of AI Risks}
To measure the global prevalence of AI risks across all six categories described in the previous section, we counted the number of articles reporting risks related to each category, and divided this number by the total number of articles across regions and categories (i.e., 16,312). Results show that prevalence of risks varies and is substantially different between categories, as shown in Figure \ref{fig:media-structure}B.

Global news coverage of AI in our sample prioritizes the coverage of societal risks, constituting approximately half of the articles in our sample (50.6\%). The reporting on \textit{Societal Risks} is primarily driven by economic risks and risks associated with potential misuse of AI for deception and manipulation, especially in politics, amounting to $\sim$ 25.3\% and 18.5\% of the overall articles in our sample. \textit{Legal and rights-related Risks} is the second most prevalent risk category in the global coverage of AI, accounting for approximately one-third (32.9\%) of the articles discussing this risk and its sub-categories. The discourse around this risk category is centered around covering privacy issues and discrimination incidents and biases caused by AI and its applications, which represent 13.6\% and 13.0\% of the total articles analyzed in our sample, respectively. The remaining four risks categories-\textit{Cognitive Risks} (14.20\%), \textit{Content Safety Risks} (8.29\%), \textit{Existential Risks} (7.24\%), and \textit{Environmental Risks} (1.86\%)-received substantially less attention in news coverage in our sample compared to \textit{Societal Risks} and \textit{Legal and Rights-related Risks}. 

Overall, the prevalence of the different risk categories irrespective of their political bias mirror the global prevalence of risks (shown in Figure \ref{fig:media-structure}B). %Specifically, Societal Risks, Legal \& Rights-related Risks are two most prevalent risk categories, followed by Cognitive Risks, Content Safety Risks, Existential Risks and Environmental Risks. 
However, an analysis of news coverage for each risk category across the spectrum of political bias reveals that political fringe media (both left and right) report more on AI risks than center, or least-biased media outlets, with a relatively similar prioritization of \textit{Societal Risks}, \textit{Legal \& Rights-related Risks}, \textit{Cognitive Risks}, and \textit{Content Safety Risks}, as illustrated in Figure \ref{fig:global-bias-prevalence}. Furthermore, we find that right-biased media outlets lead the reporting on \textit{Existential Risks} (9.1\% vs. 6.8\% for left-biased sources), but also substantively disregard \textit{Environmental Risks} (0.7\%) as compared to left bias sources (2.4\%). In contrast, least-biased media outlets consistently under report on risks compared to fringe media in nearly every risk category.

\subsubsection{Regional Prevalence of AI Risks}
Regional news coverage of AI risk categories follows the same overlying global trend of reporting on AI risks. Mainly, \textit{Societal Risks} are the most prevalent in every region, followed by \textit{Legal \& Rights-related Risks}. In contrast, the prevalence of the remaining risk categories did not exceed 20\% per region. This shows an imbalance in the nature of coverage of AI risks by news media with a clear prioritization of \textit{Societal} and \textit{Legal \& Rights-related} risks. That said, we do observe changes in the rank order of risk categories across regions (see Figure \ref{fig:prevalence-pivto} for details). We also map the political bias onto the prevalence of risks in our sample for each region. This will enable us to compare the prevalence of risk categories for each political bias group for each region, as shown in Figure \ref{fig:risks-by-political-bias} in the Appendix. In addition, this provides early insights into the potential polarization of the discourse around the risk categories within each region.

\textbf{Africa}. \textit{Societal Risks} dominate media coverage of AI in Africa in our sample, with a prevalence of 55.6\%, surpassing that of any other region for this category. In contrast, \textit{Existential Risks} received the lowest prevalence in the news reporting compared to that in other regions. Other categories followed the global trend of risk prevalence showing \textit{Legal \& Rights-related Risks} taking the second most prevalent risk category after \textit{Societal Risks} amounting to (34.6\%) of the media coverage in Africa, followed by \textit{Cognitive Risks} (15.1\%), and \textit{Content Safety Risks} (7.6\%).
Unlike other regions in our sample, media coverage of risks in Africa by least-biased sources dominates the reporting on \textit{Societal Risks} and \textit{Cognitive Risks}. However, reporting trends of other risk categories is mainly driven by left-biased sources. For instance, in comparison to left-biased media coverage, left-center media outlets show much less coverage of \textit{Cognitive Risks} (6.3\%) and \textit{Content Safety Risks} (3.1\%) in their reporting.

\textbf{Asia}. Asia's media coverage of AI was dominated by \textit{Societal Risks} (48.6\%). The most notable diversion of the coverage in Asia compared to other regions is the considerably lower focus on \textit{Legal \& Rights-related Risks} (29.5\%). Additionally, \textit{Content Safety Risks} (6.2\%) receive slightly less coverage than in other regions. Our sample of analyzed articles in Asia is politically centered for the most part, reflecting a more balanced reporting on AI and its impacts relative to other regions. However, we find media coverage in Asia by least-biased sources to report less on \textit{Societal} risks, as compared to politically-centered sources, but more on \textit{Legal \& Rights-related risks}. In addition, despite having only 25 articles from right-biased sources in our sample in Asia, these sources lead the coverage of Environmental and Existential Risks.

\textbf{Europe}. The European media coverage of AI risks is mostly aligned with the global trends shown in Figure \ref{fig:media-structure}B. \textit{Societal Risks} are  prevalent in over half of the analyzed articles (52.2\%), and the \textit{Legal \& Rights-related Risks} were present in roughly one third of the articles (34.3\%). In comparison to other regions, European media coverage has a slightly stronger focus on \textit{Cognitive Risks} (16.0\%) and \textit{Content Safety Risks} (11.5\%). Also, European media, as well as news coverage of AI from sources in the middle east, show the highest emphasis on existential threats (9.0\%) compared to that of all other regions.
The European sample consists primarily of left-center media outlets (62.5\%). In addition, there is a roughly equal share of articles by least-biased (12.4\%), right-center (10.4\%), and right-biased (12.1\%) sources, with very small representation of articles by left-biased sources (2.5\%). The European discourse shows the highest prevalence differences between political leanings of all regions. Especially left media show a strong emphasis on \textit{Legal \& Rights-related Risks} (52.1\%) and \textit{Societal Risks} (57.6\%), indicating a strong prioritization of these risks by left media in Europe. We also find stark differences in other categories like \textit{Cognitive Risks}, where left media outlets cover those issues in 20.6\% of articles, whereas least-biased media do that only in approximately one out of 10 articles (10.8\%). As for content safety risks, we find that right-biased media outlets have the highest share (13.5\%), whereas left media outlets report on those issues at a much lower rate (6.9\%). By contrast, the center-left and center-right media coverage show a comparable prevalence in each risk category.

\textbf{Middle East}. In the Middle East, news coverage shows an emphasis on \textit{Societal Risks}. Although it is still the most prioritized and prevalent category in the news coverage for that region, it is substantially less pronounced (43.0\%) compared to that of other regions. By far, \textit{Content Safety Risks} experience the highest prevalence in coverage of AI in comparison to that of other regions, reaching 15.6\%. In contrast, Cognitive Risks had the lowest prevalence across all regions amounting to 11.6\% of the articles. 
Media coverage of AI risks in our sample of articles from news outlets in the Middle East consists of left-center, right-center, and right-biased media outlets with clear dominance of right-center sources on the discourse (72\%).%, as shown in Figure \ref{fig:media-structure}A. 
In addition, least-biased and left-biased coverage of AI risks are not as prevalent as the coverage of other politically-biased domains in the region, which could be attributed to a sample bias due to the presence of a single article in the left-bias category. Also, the discourse shows signs of bipartisan emphasis across news domains on covering AI risks specifically around societal risks related issues, with left-center and right-biased media outlets showing stronger prevalence of these issues by 21.4 percentage points difference in comparison to coverage in center-right outlets. Another critical finding from this region, is the minimal concern of right-biased media coverage in our sample to cognitive and content safety risks, as reflected by the prevalence of articles on these issues, as to centered media outlets which show a considerable attention on these issues. Lastly, and consistent with other regions, \textit{Environmental Risks} are not as prioritized in media coverage as other risks. Also, issues related to these risks do not receiving much attention from news domains except left-center and right-center biased domains, amounting to 0.9\% of the coverage in that region.

\textbf{North America}. Media in North America shows consistent patterns to the global prevalence trends of AI risk categories. Again, \textit{Societal Risks} dominate the media coverage (50.9\%), \textit{Legal \& Rights-related Risks} receive the second most attention (32.9\%), followed by \textit{Cognitive Risks} (14.1\%), \textit{Content Safety Risks} (7.6\%), and \textit{Existential Risks} (6.7\%). News coverage in our sample from domains in North America is dominated by left-center media outlets, followed by right-centered, left, least biased, and right-biased outlets.

Topically, we find differences in the prevalence around \textit{Societal Risks} (58.8\%) and \textit{Legal \& Right-related Risks} (42.9\%) as they are primarily led by right-biased news media. Left-biased media, on the other hand, show the strongest emphasis on \textit{Cognitive Risks} (17.1\%) in their coverage for this category compared to media outlets of other political bias groups. Also, we generally found that least-biased media outlets have a lower prevalence of risks in their coverage, especially in \textit{Legal \& Rights-related Risks} (29.8\%) and \textit{Content Safety Risks }(5.3\%) when compared to the prevalence of these risks in left-biased and right-biased outlets.

\textbf{Oceania}. Consistent with the news coverage of AI risks in other regions, the majority of the news coverage in Oceania shows an emphasis on \textit{Societal Risks} (50.2\%). However, \textit{Legal \& Rights-related Risks} received more attention than other regions amounting to 38.3\% of the news coverage of AI in our sample. This is the hightest prevalence for this category across all regions. Likewise, \textit{Cognitive Risks} received the hightest share of media attention in Oceania compared to all other regions (16.4\%). 
For Oceania, our sample consists mostly of left-center outlets, followed by least biased, and right center outlets. Right-biased and left-biased news outlets are only marginally present. Similar to North America, we find signs of potential bipartisan emphasis in the discourse around AI risks in Oceania. Specifically, right-biased news outlets show a strong emphasis on \textit{Societal Risks}, which represents 61.5\% of all articles published in the region. As for the legal \& rights-related risks, least-biased media outlets, show a strong prevalence of these risks as compared to the coverage by other politically biased outlets. Notably, we also found that right-biased news sources in this region overly emphasize existential risks. Specifically, 23.1\% of the articles published by right-biased sources in this region cover issues related to \textit{Existential Risks}, making it the highest share among all regions and biases in this category.

%% file: discussion.tex
In this research, we analyze a broad and diverse sample of global news media spanning 27 countries to unlock insights about AI risks and harms reported in media outlets globally across six regions: Africa, Asia, Europe, Middle East, North America, and Oceania. Although our sample represents media coverage from six regions worldwide, it is subject to certain limitations and biases. First, our sample relies primarily on news domains that are ``freely available" (i.e., not behind paywalls) and report on AI risks in the English language, which does not capture the full discourse around AI in languages native to several of the regions we are analyzing. In addition, it should be recognized that the reliance on English as a language in our sample for news reporting, especially for countries in the Global South, is influenced by historical and cultural ties to the British Commonwealth. This colonial legacy may influence how these nations view and interpret the economic, political, and cultural risks associated with AI, potentially biasing how AI is reported on in these countries. Thus, future work should explore conducting evaluations of AI risks in languages beyond English. Accordingly, the results of this work and our interpretations of them are limited to publicly accessible news domains, the quality of information in these domains, and the limited representation of these sources to the views of the regions they are publishing from, particularly in non-English speaking regions. Second, although our sample includes AI risks from media coverage in 27 countries, it does not include AI risks reported in Latin America. This highlights a gap in our efforts to achieve a more comprehensive and global understanding of the prevalence of the various types of AI risks in news media.

The analysis of the news coverage of AI in our sample shows a global prioritization of news media, as reflected in the prevalence of these risks in our sample, in covering societal risks, followed by legal and rights-related risks, cognitive risks, content safety risks, existential risks, and environmental risks. One possible explanation for the large degree of global convergence on covering these categories of risks could be that there may be external forces driving the anglophone media system towards isomorphism \cite{becker2024policies}, including aspects such as tech-industry influence of media coverage \cite{brennen2021balancing}, or media norms such as newsworthiness which could serve to homogenize coverage.  We also observe that regional coverage of AI in the news, per the prevalence of AI risks in each region in our sample, seem to align with the overall global trends of AI risks, with some variance between regions. For instance, in the Middle East content safety risks were the third most prevalent whereas other regions had it as the fourth most prevalent impact reported, and Europe having existential risks be the fourth most prevalent whereas others had it as the fifth (see Figure \ref{fig:prevalence-pivto}). 

The emphasis of news media on societal issues in our sample, especially those focusing on economic and political risks, could be reflecting the ongoing global narrative of investing in AI to avoid missing out on opportunities for economic growth (i.e., the global AI race) \cite{cave_ai_2018, crawford_atlas_2021}. We also speculate that the emergence of legal \& rights-related risks as the second most prevalent risk in media coverage of AI could be partly due to the growing academic and regulatory discourse around rights-related question related to privacy and copyright, but also the protection of fundamental rights \cite{orwat_normative_2024,ohchr2023}. Moreover, we find it surprising that cognitive risks and content-safety risks receive substantially less attention in media reporting, despite involving severe risks to individuals \cite{barnett_simulating_2024-1}. For example, cognitive risks include mental \& emotional risks, while content safety risks include issues such as hate speech and pornographic deepfakes, which cause tremendous harm to those affected by them. We also observe that existential risks are a small but not entirely insignificant part of the reporting of AI in news media. News media coverage of AI appears to be more concerned about present tangible risks that are being realized or manifested in society, more so than on what public figures may emphasize regarding the potential dangers of AI and super-intelligence on the human race. This is also consistent with prior research providing little evidence that news media has succumbed to the rhetoric of a doomsday scenario of AI \cite{cave_portrayals_2018, gilardi_we_2024, allaham_evaluating_2024}. Lastly, the nearly non-existent media coverage on environmental risks is perplexing in light of the ongoing politicized and polarized discussion around climate change policies \cite{chinn2020politicization}. This suggests that climate debates have not gained substantial traction in the discourse on AI, despite the significant resources involved and the potential environmental impacts highlighted by our research. More strikingly, a majority of the present risk assessments tend to overlook the environmental risks of AI \cite{crawford_atlas_2021, zeng_ai_2024, stahl_systematic_2023}, except a few \cite{solaiman_evaluating_2023, allaham_evaluating_2024, uuk_taxonomy_nodate}, as if the infrastructure required for AI systems is only marginally connected to environmental issues.

When inspecting the potential influence of political bias of the news domains reporting on AI, we mainly find that fringe media (left- and right-biased sources) tend to report more on risks - especially in comparison to the least-biased outlets. This perhaps makes sense given that underlying every risk is a value judgment about what matters and to whom. Also, we observe instances of potential politicization of the discourse around AI such as the one surrounding legal and rights-related risks in left-biased news media in Europe, echoing results of some scholars \cite{brennen_industry-led_2018, vergeer_artificial_2020}. This could be attributed to the discussion around the introduction of the EU AI Act, where fundamental rights and societal values played an important role \cite{orwat_normative_2024}. We find similar patterns in other regions such as in Oceania where right-biased media focus more on societal and existential risks, and North America where the prevalence of societal risks and legal and rights-related risks are more emphasized by right-wing media outlets. As for Africa, we found higher differences in political bias in reporting about individual risks (cognitive impacts and content safety risks) indicating a potential polarized debate for these dimensions. Explaining these differences is beyond the scope of this descriptive paper, as it requires expertise in each of the six regions to interpret the structural and contextual differences in AI reporting especially if the discourse is being shaped by local sources to these regions that are politically-biased. Therefore, we encourage future research to look into explaining the influence of political bias of news outlets on the reporting and framing of AI risks, as this may contribute to mitigating future polarizing debates around AI risks, such as economic risks \cite{huang2024uncovering}, that could influence (or weaponize) the public perception of AI during the current and future development of AI regulations and policies.

%% file: conclusion.tex
This work highlights the importance of incorporating global perspectives, both regionally and politically, when assessing the broader impacts of AI risks and harms. One source rich with such diverse perspectives, is news media. By analyzing news coverage of AI across six regions around the world, we find a global skew towards prioritizing societal risks and legal and rights-related risks, in comparison to other risk categories, in the reporting of AI in our sample from news media. Our research presents stakeholders, such as AI developers and policymakers, with insights into the categories of AI risks prioritized globally and within each of the six regions. Further, by evaluating the political bias of the news sources contributing to this coverage, we highlight the potential influence of the political bias of news outlets on the coverage of AI risks. If such bias is left unaccounted for in future evaluations of AI and its impacts, it may influence the public perception of AI--who is key stakeholder in shaping the current and future development of AI regulations and policies. Thus, hindering any progress towards shaping more inclusive and globally informed AI governance policies and regulations.

%% file: appendix.tex
\subsection{AI-relevant Keywords}\label{a3}
The set of keywords used to probe the news media for articles on AI sourced from \cite{allaham2024towards}:\\
A.I., Artificial Intelligence, Automated Decision Making, Automated System, Autonomous Driving System, Autonomous Vehicles, Autonomous Weapon, Chat Bot, Chatbot, ChatGPT, Computer Vision, Deep Learning, Deepfake, Driverless Car, Facial Recognition, General Artificial Intelligence, Generative AI, GPT, Image Generator, Intelligence Software, Intelligent Machine, Intelligent System, Language Model, Large Language Model, LLMs, Machine Intelligence, Machine Learning, Machine Translation, Natural Language API, Natural Language Processing, Neural Net, Neural Network, Predictive Policing, Reinforcement Learning, Self-Driving Car, Speech Recognition, Stable Diffusion, Synthetic Media, Virtual Reality, Weapons System.
\newpage
\include{table-dist-of-articles-per-country}
\include{table-distribution-of-human-annotated-impacts}
\newpage
\include{table-classification-report-on-impact-categories-at-level-1}
\newpage
\include{appendix-prevalence-of-risks-per-region-stratified-by-political-bias}
\newpage
\include{appendix-prompts}
\newpage
\include{appendix-definition-of-annotated-risk-categories}
\newpage
\input{table-distribution-of-llm-annotated-categories}
\newpage
\input{appendix-clustering-method}
\begin{figure*}
\centering
\includegraphics[width=0.75\textwidth]{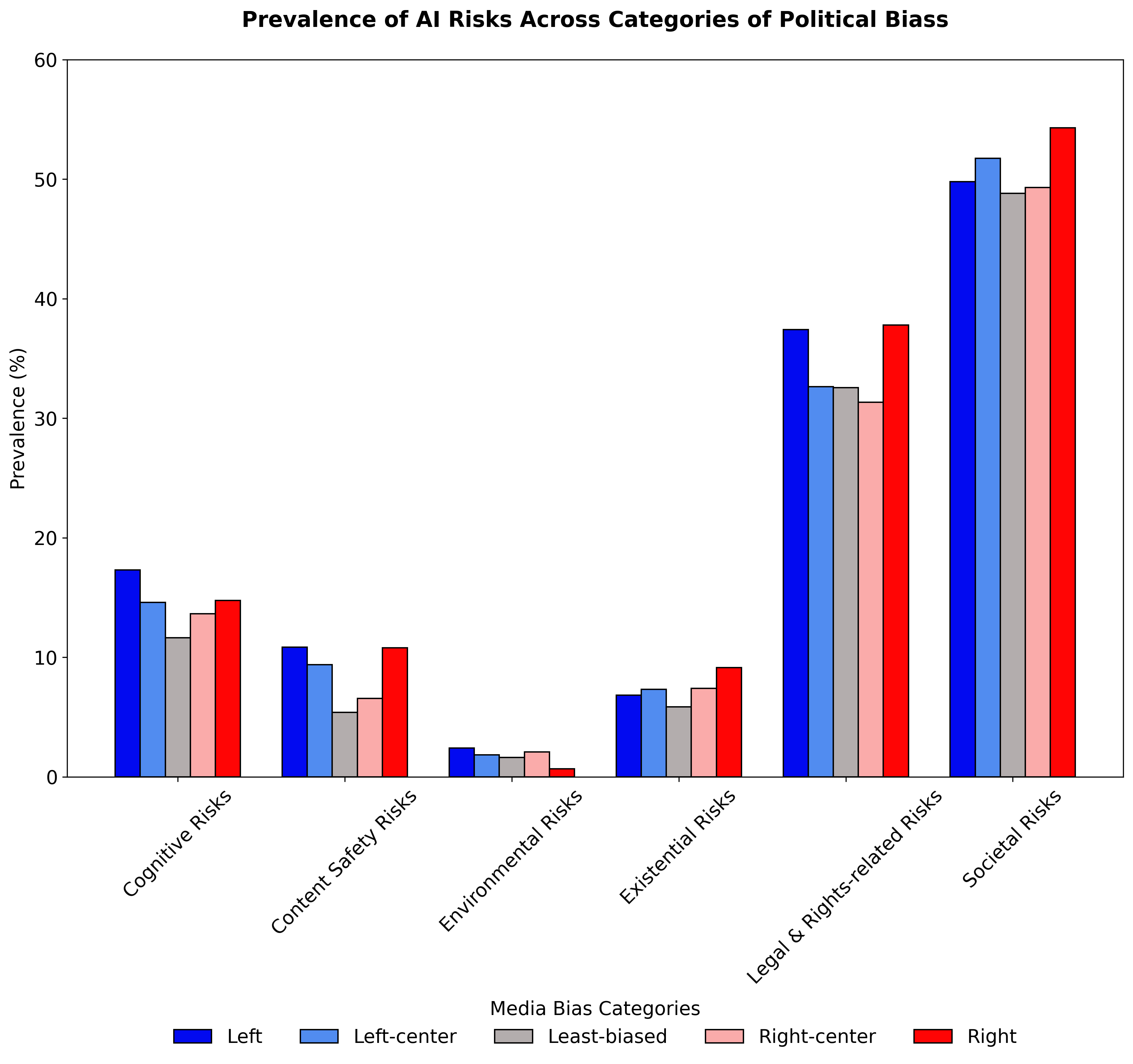} % Path to the image file
\caption{Global prevalence of AI risks in our sample from news media coverage across the categories of political bias.}
\label{fig:global-bias-prevalence}
\end{figure*}

%% file: table-dist-of-articles-per-country.tex
\onecolumn
\subsection{Distribution of news articles in our sample}
\begin{table}[H]
\centering
\small
\setlength{\tabcolsep}{6pt}
\renewcommand{\arraystretch}{1.2}
\begin{tabularx}{\textwidth}{>{\raggedright\arraybackslash}p{3cm} >{\raggedright\arraybackslash}p{4cm} >{\centering\arraybackslash}p{3cm} >{\centering\arraybackslash}p{3cm}}
\toprule
\textbf{Region} & \textbf{Country} & \textbf{Articles} & \textbf{Proportion (\%)} \\
\midrule
\textbf{Africa} & Nigeria          & 936   & 2.184  \\
                & South Africa     & 484   & 1.129  \\
                & Kenya            & 235   & 0.548  \\
                & Uganda           & 38    & 0.089  \\
\midrule
\textbf{Asia}   & India            & 4957  & 11.567 \\
                & Japan            & 1560  & 3.640  \\
                & Philippines      & 1230  & 2.870  \\
                & Pakistan         & 1245  & 2.905  \\
                & China            & 549   & 1.281  \\
                & Singapore        & 468   & 1.092  \\
                & Malaysia         & 166   & 0.387  \\
                & Russia           & 196   & 0.457  \\
                & Taiwan           & 42    & 0.098  \\
\midrule
\textbf{Europe} & United Kingdom   & 4846  & 11.308 \\
                & Ireland          & 1320  & 3.080  \\
                & Malta            & 359   & 0.838  \\
                & Germany          & 253   & 0.590  \\
                & France           & 217   & 0.506  \\
                & Bulgaria         & 54    & 0.126  \\
                & Italy            & 26    & 0.061  \\
                & Iceland          & 6     & 0.014  \\
\midrule
\textbf{Middle East} & Israel      & 1373  & 3.204  \\
                & Turkey           & 115   & 0.268  \\
\midrule
\textbf{North America} & United States & 18399 & 42.935 \\
                & Canada           & 1923  & 4.487  \\
\midrule
\textbf{Oceania} & Australia        & 1193  & 2.784  \\
                 & New Zealand      & 649   & 1.51  \\
\bottomrule
\end{tabularx}
\caption{Distribution of the number of scraped articles and their prevalence (out of 42,853 articles) in our corpus per region and country.}
\label{tab:tab3}
\end{table}

%% file: table-distribution-of-human-annotated-impacts.tex
\onecolumn
\subsection{Distribution of AI Risks categories annotated by two authors}
\begin{table}[H]
\centering
\small
\captionsetup{width=\columnwidth}
\setlength{\tabcolsep}{4pt} % Adjust the horizontal padding
\renewcommand{\arraystretch}{1.2} % Adjust the row height
\begin{tabularx}{\textwidth}{>{\raggedright\arraybackslash}X>{\centering\arraybackslash}X>{\centering\arraybackslash}X}
\toprule
\textbf{Class} & \textbf{Count} & \textbf{\% Prevalence} \\
\midrule
Economic Risks & 140 & 13.21 \\
Deception/Manipulation Risks & 100 & 9.43 \\
No Impact & 97 & 9.15 \\
Privacy Risks & 72 & 6.79 \\
AI Governance Risks$^\ast$ & 65 & 6.13 \\
Operational Misuses Risks & 57 & 5.38 \\
Information Risks$^\ast$ & 54 & 5.09 \\
Fundamental Rights & 50 & 4.72 \\
Existential Risks$^\ast$ & 38 & 3.58 \\
Discrimination/Bias Risks & 37 & 3.49 \\
Political Usage Risks & 37 & 3.49 \\
User Experience Risks$^\ast$ & 36 & 3.40 \\
Security Risks & 29 & 2.74 \\
Ethical Risks$^\ast$ & 28 & 2.64 \\
Humanness Risks $^\ast$ & 27 & 2.55 \\
Structure/Power Risks$^\ast$ & 22 & 2.08 \\
Mental \& Emotional Risks$^\ast$ & 20 & 1.89 \\
Performance Risks$^\ast$ & 20 & 1.89 \\
Environmental Risk$^\ast$ & 18 & 1.70 \\
Safety Risks$^\ast$ & 17 & 1.60 \\
Hate/Toxicity Risks & 16 & 1.51 \\
Authoritarian Use of AI Risks$^\ast$ & 15 & 1.42 \\
Violence \& Extremism Risks & 12 & 1.13 \\
Criminal Activities Risks & 11 & 1.04 \\
Over-reliance Risks$^\ast$ & 10 & 0.94 \\
Technology Adoption Risks$^\ast$ & 10 & 0.94 \\
Sexual Content Risks & 7 & 0.66 \\
Disruption of Service Risks$^\ast$ & 6 & 0.57 \\
Media Risks$^\ast$ & 5 & 0.47 \\
Defamation Risks & 2 & 0.19 \\
Child Harm Risks & 2 & 0.19 \\
\bottomrule
\end{tabularx}
\caption{Distribution of risk categories in a sample of 1,060 summarized impacts that were annotated by two authors as described in section \ref{4.2}. All risk categories appended with an asterisk ($^\ast$) are novel risk categories that are not part of the AIR-taxonomy \cite{zeng_ai_2024} at any level.}
\label{tab:tab1}
\end{table}

%% file: table-classification-report-on-impact-categories-at-level-1.tex
\onecolumn
\begin{table}[ht]
\centering
\small % Adjust font size as needed
\captionsetup{width=\textwidth}
\setlength{\tabcolsep}{5pt} % Adjust column padding
\renewcommand{\arraystretch}{1.2} % Adjust row height
\captionsetup{width=0.9\textwidth} % Set the caption width
\begin{tabularx}{0.9\textwidth}{>{\raggedright\arraybackslash}p{3.5cm} 
>{\centering\arraybackslash}p{2cm} 
>{\centering\arraybackslash}p{2cm} 
>{\centering\arraybackslash}p{2cm} 
>{\centering\arraybackslash}p{2cm}}
\toprule
\textbf{Category} & \textbf{Precision} & \textbf{Recall} & \textbf{F1-Score} & \textbf{Support} \\
\midrule
AI Governance Risks*                  & 0.46 & 0.54 & 0.50 & 65 \\
Cognitive Risks*               & 0.65 & 0.77 & 0.70 & 57 \\
Content Safety Risks          & 0.58 & 0.70 & 0.63 & 37 \\
Environmental Risks*                  & 1.00 & 1.00 & 1.00 & 18 \\
Ethical Risk*                 & 0.47 & 0.57 & 0.52 & 28 \\
Existential Risks*          & 0.75 & 0.95 & 0.84 & 38 \\
Information Risks*              & 0.60 & 0.56 & 0.58 & 54 \\
Legal \& Rights-related Risks & 0.67 & 0.77 & 0.72 & 170 \\
No Impact                      & 0.83 & 0.05 & 0.10 & 97 \\
Societal Risks                 & 0.69 & 0.82 & 0.75 & 299 \\
Structure/Power*                 & 0.43 & 0.27 & 0.33 & 22 \\
Sytem \& Operational Risks & 0.54 & 0.49 & 0.51 & 129 \\
Usability Risks*               & 0.69 & 0.48 & 0.56 & 46 \\
\midrule
\textbf{Accuracy}               &       &       & 0.64 & 1060 \\
\textbf{Macro Avg}              & 0.64 & 0.61 & 0.60 & 1060 \\
\textbf{Weighted Avg}           & 0.65 & 0.64 & 0.61 & 1060 \\
\bottomrule
\end{tabularx}
\caption{Classification report showing the performance of GPT-4o in annotating negative impacts in a sample of 1,060 annotated impacts by two annotators, as described in section \ref{4.2}. Risk categories marked with an asterisk (*) represent the emerging categories in our sample that we attempted to align with the AIR-taxonomy, as outlined in Section \ref{5.1}.}
\label{tab:tab2}
\end{table}

%% file: appendix-prevalence-of-risks-per-region-stratified-by-political-bias.tex
\begin{figure*}
\centering
\includegraphics[width=0.8\textwidth]{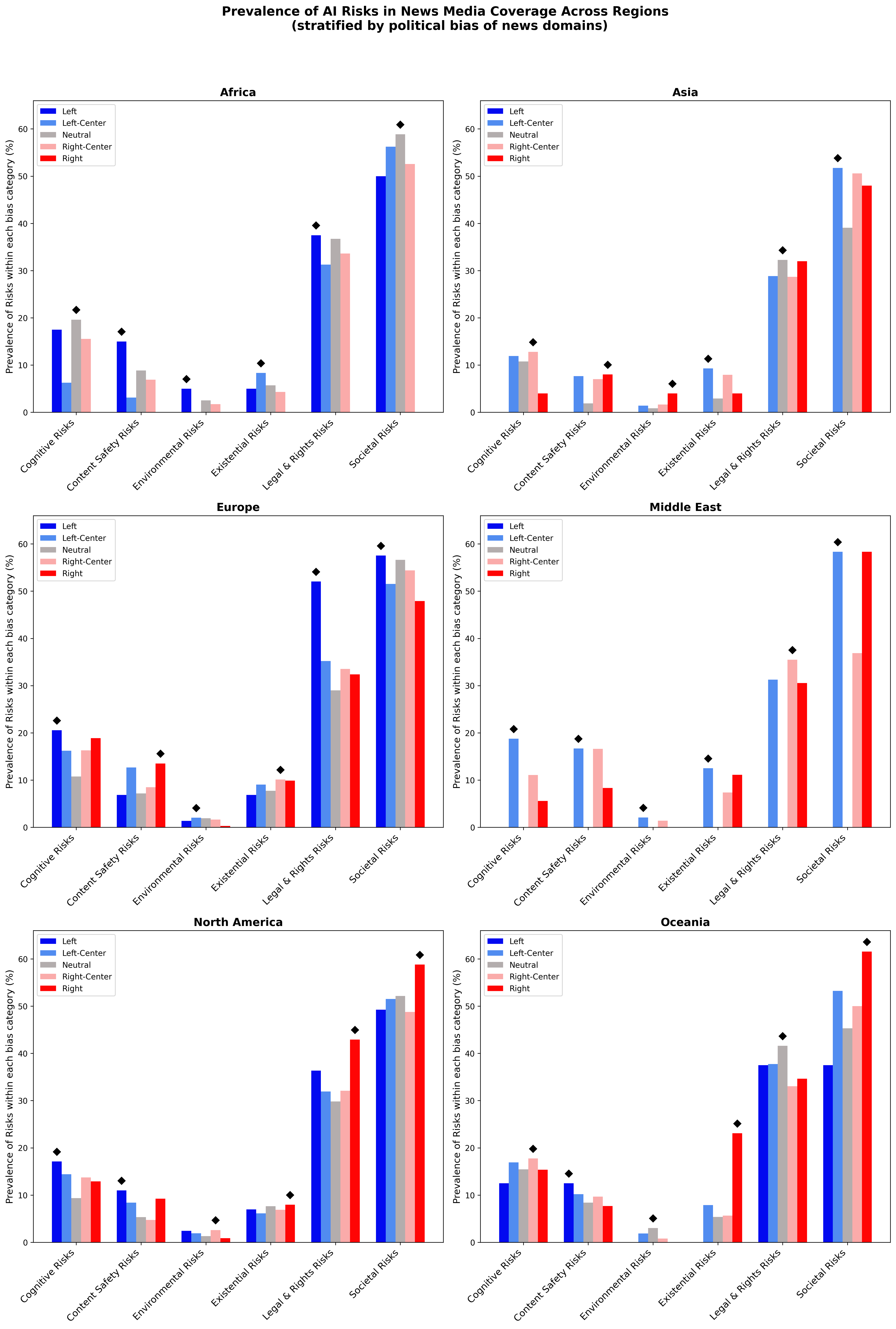} % Path to the image file
\caption{Prevalence of AI risks in our sample from news media coverage across six different regions stratified by the political bias of news domains as rated by Media Bias Fact Check (MBFC) and elaborated on in Section \ref{3.3}. We marked the political bias leading the discourse for risks in each region by a diamond.}
\label{fig:risks-by-political-bias}
\end{figure*}

%% file: appendix-prompts.tex
\twocolumn
\subsection{Prompt for filtering articles by content}\label{p1}
\begin{lstlisting}[language=Python, caption={Prompt to classify whether an article describes an impact of an AI system based on the conceptual definition provided in the prompt.},  numbers=none, breaklines=true]
### Context
{article_text}

### Definition
An impact refers to an effect, consequence, or outcome of an AI system (i.e., model or application) that positively or negatively affects individuals, organizations, communities, or society.

### Task
Based on the definition of an impact of an AI system provided to you, does the article above cover or describe at least one impact of an AI system? Answer Yes or No.

DO NOT explain yourself.
\end{lstlisting}

\subsection{Prompt for summarizing negative impacts from articles}\label{p2}
\begin{lstlisting}[language=Python, caption={Prompt to summarize the negative impacts of AI in an article based on the provided operational definition of a negative impact.},  numbers=none, breaklines=true]

Summarize ALL negative impacts that are ONLY relevant to the AI system or model described in the context provided to you. A negative impact refers to the set of risks or harms that have or may affect individuals, organizations, or communities in society, as a result of an AI system, or its use. Each summarized impacts must be 1 sentence long and must have sufficient details that are grounded in the context provided to you.

Format your response in a jsonl format {"negative_impacts": [list of impacts]}. If not impacts are present, output {"negative_impacts":[]}.

DO NOT make up details for impacts.
DO NOT interpret any details, share any highlights, or draw conclusions.
ONLY stick to the context provided to you.
DO NOT provide any other details in the answer besides the jsonl content.

### Context:
{article_text}
\end{lstlisting}

\subsection{Prompt for annotating negative impacts in corpus of articles}\label{p3}
\begin{lstlisting}[language=Python, caption={Prompt to scale up the annotation process of negative impacts of AI using GPT-4o as described in section \ref{4.2}.},  numbers=none, breaklines=true]

Task: Analyze the negative impact statements of AI technologies. A negative impact refers to the set of risks or harms that have or may affect individuals, organizations, or communities in society, as a result of an AI system, or its use. Each impact statement should be evaluated and categorized into ONLY one of the following 32 categories. For each impact statement listed, assign exactly one label that corresponds to its most appropriate classification in the same format as the original list. 
Categories :

1. ai_governance: negative impacts associated with governance policies and regulations related to the development, deployment, licensing, or moderation of AI technologies.

2. ai_incompetence: risks resulting from limitations and malfunctions of AI technologies that impact their performance.

3. authoritative_use_of_ai: risks associated with the potential misuse of artificial intelligence by governments in ways that may support authoritarian practices that violate human rights and civil liberties.

4. criminal_activities: risks associated with the misuse of AI technologies for online crimes such as cybercrimes or cyberattacks.

5. deception/manipulation: risks associated with the use of AI technologies for fraud, dishonest activities, sowing divisions, or misrepresenting individuals to influence or alter perceptions, behaviors, or mislead individuals or society.

6. discrimination/bias: risks of AI technologies generating outputs based on protected characteristics that result in unequal treatment or representation of individuals or social groups.

7. disruption_of_service: risks related to disruptions or reductions in the accessibility, availability, and functionality of AI systems.

8. economic_harm: risks posed by AI technologies to financial systems, labor market, and trading dynamics.

9. environmental: environmental and ecological risks arising from the energy consumption and resource intensive process required for the development, deployment, and operation of AI technologies.

10. ethical_impact: challenges related to inequalities in access to AI technologies, or in their development, deployment, and use, with a particular focus on issues of fairness, accountability, and transparency.

11. existential_threats: risks related to potential inequalities in accessing AI technologies, their development, deployment, and use, with a particular focus on issues of fairness, accountability, and transparency.

12. fundamental_rights: risks posed by AI technologies related to violating individual freedoms and rights, including freedom of expression and intellectual property.

13. information_risks: risks associated with AI hallucinations, including the generation of inaccurate information, low-quality AI-generated content, and fabrication of information by AI technologies.

14. hate/toxicity: risks associated with AI-generated content amplifying or spreading hateful, abusive, or offensive content.

15. humanness: risks related to the loss or diminishment of human qualities, such as creativity, emotional depth, and authentic interpersonal connections, due to AI technologies.

16. media_impacts: risks of AI technologies on the independence, integrity, and reliability of media and journalism.

17. mental_&_emotional: risks related to the psychological well-being and emotional health of individuals using or interacting with AI technologies.

18. operational_misuses: risks associated with the misuse of AI technologies in critical and highly regulated applications, such as unsafe autonomous operations, unreliable legal or military advice, or automated decision-making.

19. over-reliance: risks arising from over-relying on AI technologies in contexts that results in undermining human judgment, critical thinking, and decision-making.

20. political_useage: risks associated with the use of AI technologies to spread misinformation or disinformation, influence elections or politics, undermine democratic integrity, or disrupt social order.

21. privacy: risks related to unauthorized access, use, or disclosure of users' data and personal information.

22. safety_risks: risks of harming or endangering individuals' lives or safety arising from the malfunction or misuse of AI technologies.

23. security_risks: risks related to the threats and exploitation of vulnerabilities that compromise the confidentiality, integrity, or availability of AI technologies.

24. sexual_content: risks related to the non-consensual creation, distribution, or misuse of sexually explicit material or pornography using AI technologies.

25. structure/power: risks related to the concentration of power and AI resources or technologies among a few entities or governments, and its consequences on competition, collaboration, innovation, and safety of AI technologies.

26. technology_adoption: risks related to the adoption of AI technologies due to integration and usability challenges, or due to barriers faced by organizations and individuals in adopting AI technologies into their work.

27. user_experience: risks and issues that undermine the satisfaction satisfaction, trust, and interaction of the end-user with AI technologies.

28. violence_&_extremism: risks pertaining to the use of AI technologies or AI-generated content to incites violence, promotes extremist ideologies, or enables harmful activities such as weapon development or using AI in warfare.

29. defamation: risks involving reputational harms to individuals or organizations through AI-generated false or misleading statements, images, or representations.

30. child_harm: risks related to the misuse of AI technologies to harm or exploit children.

31. no_impact: refers to general statements that do not highlight potential or direct negative consequences, risks, or harms of AI systems.

32. other: Any risks or harms that do not fit into the above categories.

Output Format : Present the classified categories without any numbers and clean from whitespace. The categories should be selected from one of the above 33 categories.

Note: Ensure that each impact statement is classified under only one of the aforementioned 32 categories and that there is one
classification corresponding to each impact statement in the input. Do not generate any other text or include any additional details.
Do not make up categories.

Impact:
{summarized_impact}
\end{lstlisting}

%% file: appendix-definition-of-annotated-risk-categories.tex
\onecolumn
\subsection{Definitions of Impact Categories}\label{definitions-of-impacts}
Below are the definitions of the impact categories, grouped at Level-1 of the AIR-taxonomy, resulted from annotating a sample of 1,060 summaries of negative impacts by two authors as outlined in Section \ref{4.2}. Novel categories that are not part of the AIR-taxonomy are formatted in bold.

\subsection*{1. Societal Risks}
\begin{itemize}
     \item \textit{Economic Risks}: risks posed by AI technologies to financial systems, labor market (incl. job safety), and trading dynamics.
    \item \textbf{\textit{Authoritarian Use of AI Risks}}: risks associated with the potential misuse of artificial intelligence by governments in ways that may support authoritarian practices that violate human rights and civil liberties.
    \item \textit{Deception/Manipulation Risks}: risks associated with the use of AI technologies for fraud, dishonest activities, sowing divisions, or misrepresenting individuals to influence or alter perceptions, behaviors, or mislead individuals or society.
    \item \textit{Political Usage Risks}: risks associated with the use of AI technologies to spread misinformation or disinformation, influence elections or politics, undermine democratic integrity, or disrupt social order.
    \item \textit{Defamation Risks}: risks involving reputational harms to individuals or organizations through AI-generated false or misleading statements, images, or representations.
    \item \textbf{\textit{Media Risks}}: risks of AI technologies on the independence, integrity, and reliability of media and journalism.
\end{itemize}

\subsection*{2. Legal and Rights-related Risks}
\begin{itemize}
    \item \textit{Privacy Risks}: risks related to unauthorized access, use, or disclosure of users' data and personal information.
    \item \textit{Fundamental Rights Risks}: risks posed by AI technologies related to violating individual freedoms and rights, including freedom of expression and intellectual property.
    \item \textit{Discrimination/Bias Risks}: risks of AI technologies generating outputs based on protected characteristics that result in unequal treatment or representation of individuals or social groups.
    \item \textit{Criminal Activities Risks}: risks associated with the misuse of AI technologies for online crimes such as cybercrimes or cyberattacks. 
\end{itemize}

\subsection*{3. Content Safety Risks}
\begin{itemize}
    \item \textit{Child Harm Risks}: risks related to the misuse of AI technologies to harm or exploit children.
    \item \textit{Hate/Toxicity Risks}: risks associated with AI-generated content amplifying or spreading hateful, abusive, or offensive content.
    \item \textit{Sexual Content Risks}: risks related to the non-consensual creation, distribution, or misuse of sexually explicit material or pornography using AI technologies.
    \item \textit{Violence \& Extremism Risks}: risks pertaining to the use of AI technologies or AI-generated content to incite violence, promote extremist ideologies, or enable harmful activities such as weapon development or using AI in warfare.
    \item \textit{Hate/Toxicity Risks}: risks associated with AI-generated content amplifying or spreading hateful, abusive, or offensive content.
\end{itemize}

\subsection*{4. Cognitive Risks}
\begin{itemize}
    \item \textbf{\textit{Mental \& Emotional Risks}}: risks related to the psychological well-being and emotional health of individuals using or interacting with AI technologies.
    \item \textbf{\textit{Humanness Risks}}: risks related to the loss or diminishment of human qualities, such as creativity, emotional depth, and authentic interpersonal connections, due to AI technologies.
    \item \textbf{\textit{Over-reliance Risks}}: risks arising from over-relying on AI technologies in contexts that result in undermining human judgment, critical thinking, and decision-making.     
\end{itemize}

\subsection*{5. Existential Risks}
\begin{itemize}
    \item \textbf{\textit{Existential Risks}}: risks concerning the potential extinction of humanity or collapse of human civilization due to artificial intelligence.
\end{itemize}

\subsection*{6. Environmental Risks}
\begin{itemize}
    \item \textbf{\textit{Environmental Risks}}: environmental and ecological risks arising from the energy consumption and resource intensive processes required for the development, deployment, and operation of AI technologies.
\end{itemize}

The remaining categories that emerged from the annotation, but were excluded from further analysis (novel categories that are not in the AIR-taxonomy are formatted in bold):

\begin{itemize}
    \item \textbf{\textit{AI Governance Risks}}: negative impacts resulting from governance policies and regulations related to the development, deployment, licensing, or moderation of AI technologies.
    
    \item \textit{Operational Misuse Risks}: risks associated with the misuse of AI technologies in critical and highly regulated applications, such as unsafe autonomous operations (e.g., transportation, energy supply, military), unreliable advice in heavily regulated areas (e.g., legal, medical), or automated decision-making (e.g., social scoring, profiling).
    
    \item \textbf{\textit{Information Risks}}: risks associated with AI hallucinations, including the generation of inaccurate information, low-quality AI-generated content, and fabrication of information by AI technologies.
    
    \item \textbf{\textit{User Experience Risks}}: risks and issues that undermine the satisfaction, trust, and interaction of the end-user with AI technologies.
    
    \item \textit{Security Risks}: risks related to the threats and exploitation of vulnerabilities that compromise the confidentiality, integrity, or availability of AI technologies.
    
    \item \textbf{\textit{Ethical Risks}}: risks related to potential inequalities in accessing AI technologies, their development, deployment, and use, with a particular focus on issues of fairness, accountability, and transparency.
    
    \item \textbf{\textit{Structure/Power Risks}}: risks related to the concentration of power and AI resources or technologies among a few entities or governments, and its consequences on competition, collaboration, innovation, and safety of AI technologies.
    
    \item \textbf{\textit{Performance Risks}}: risks resulting from limitations and malfunctions of AI technologies that impact their performance.
    
    \item \textbf{\textit{Safety Risks}} risks of harming or endangering individuals' lives or safety arising from the malfunction or misuse of AI technologies.
    
    \item \textit{Criminal Activities Risks}: risks associated with the misuse of AI technologies for online crimes such as cybercrimes or cyberattacks.
    
    \item \textbf{\textit{Technology Adoption Risks}}: risks related to the adoption of AI technologies due to integration and usability challenges, or due to barriers faced by organizations and individuals in adopting AI technologies into their work.
        
    \item \textbf{\textit{Disruption of Service Risks}}: risks related to disruptions or reductions in the accessibility, availability, and functionality of AI systems.
    
    \item \textit{Defamation Risks}: risks involving reputational harms to individuals or organizations through AI-generated false or misleading statements, images, or representations.
\end{itemize}

%% file: table-distribution-of-llm-annotated-categories.tex
\begin{table*}[ht]
\centering
\small
\setlength{\tabcolsep}{2pt}
\renewcommand{\arraystretch}{0.8}
\begin{tabularx}{0.6\textwidth}{>{\raggedright\arraybackslash}p{4cm} >{\centering\arraybackslash}p{3cm} >{\centering\arraybackslash}p{2cm}}
\toprule
\textbf{Category} & \textbf{Number of Impacts} & \textbf{Proportion (\%)} \\
\midrule
Societal Risks                  & 14,824 & 31.057 \\
Legal and Rights-Related Risks  & 9,169  & 19.210 \\
System and Operational Risks    & 6,766  & 14.175 \\
Cognitive Risks*                & 3,345  & 7.008  \\
AI Governance Risks*                   & 2,926  & 6.130  \\
Information Risks*               & 2,464  & 5.162  \\
Content Safety Risks            & 2,353  & 4.930  \\
Usability Risks*                & 1,720  & 3.604  \\
Existential Risks*             & 1,647  & 3.451  \\
Ethical Risks*                  & 1,107  & 2.319  \\
Environmental Risks*                   & 560    & 1.173  \\
Structure/Power Risks*                 & 466    & 0.976  \\
\bottomrule
\end{tabularx}
\caption{Distribution of risk categories annotated by the LLM over our corpus. Categories marked with an asterisk (*) represent the eight categories we suggested to include in the AIR-taxonomy, as described in Section \ref{5.1}.}
\label{tab:tab4}
\end{table*}

%% file: appendix-clustering-method.tex
\subsection{Clustering negative impacts from news media}\label{clustering-method}
As described in Section \ref{4.2}, we tried clustering the negative impacts by mapping each summary of a single negative impact to a 384-dimensional dense vector (i.e., embeddings) using \texttt{all-MiniLM-L6-v2} sentence-transformer \cite{reimers-2019-sentence-bert}. Then, we applied the UMAP dimensionality reduction algorithm to have a low dimensional representation of the embeddings so the clustering of a large number of negative impacts would be computationally more efficient \cite{mcinnes2018umap}. Finally, we clustered the negative impacts using HDBSCAN \cite{hdbscan2025} resulting in 565 clusters. The unexpected large number of clusters led us to focus on selecting the largest 212 clusters based on the number of clustered impacts that collectively encompass 80\% of the 47,731 negative impacts. To ensure the intra-cluster consistency with respect to the types of impacts clustered, we sampled two impacts from each of the 212 clusters and qualitatively checked whether they consistently describe a similar impact. At this stage, we observed that the impact statements within each cluster refer to different types of impacts. However, the contextual use or application of the AI in these clusters is somewhat consistent. This prompted us to switch our annotation process of grouping negative impacts from clustering the negative impacts to manually annotating a sample of these impacts, and then scaling the annotation process using an LLM, as described in Section \ref{4.2}.

%% file: main.bbl

\begin{thebibliography}{111}

%%% the Web address.

\ifx \showCODEN    \undefined \def \showCODEN     #1{\unskip}     \fi
\ifx \showDOI      \undefined \def \showDOI       #1{#1}\fi
\ifx \showISBNx    \undefined \def \showISBNx     #1{\unskip}     \fi
\ifx \showISBNxiii \undefined \def \showISBNxiii  #1{\unskip}     \fi
\ifx \showISSN     \undefined \def \showISSN      #1{\unskip}     \fi
\ifx \showLCCN     \undefined \def \showLCCN      #1{\unskip}     \fi
\ifx \shownote     \undefined \def \shownote      #1{#1}          \fi
\ifx \showarticletitle \undefined \def \showarticletitle #1{#1}   \fi
\ifx \showURL      \undefined \def \showURL       {\relax}        \fi
% invisible to TeX
\providecommand\bibfield[2]{#2}
\providecommand\bibinfo[2]{#2}
\providecommand\natexlab[1]{#1}
\providecommand\showeprint[2][]{arXiv:#2}

\bibitem[Allaham and Diakopoulos(2024)]%
        {allaham_evaluating_2024}
\bibfield{author}{\bibinfo{person}{Mowafak Allaham} {and} \bibinfo{person}{Nicholas Diakopoulos}.} \bibinfo{year}{2024}\natexlab{}.
\newblock \bibinfo{title}{Evaluating the {Capabilities} of {LLMs} for {Supporting} {Anticipatory} {Impact} {Assessment}}.
\newblock
\newblock
\urldef\tempurl%
\url{https://doi.org/10.48550/ARXIV.2401.18028}
\showDOI{\tempurl}
\newblock
\shownote{Version Number: 2}.


\bibitem[Allaham et~al\mbox{.}(2024)]%
        {allaham2024towards}
\bibfield{author}{\bibinfo{person}{Mowafak Allaham}, \bibinfo{person}{Kimon Kieslich}, {and} \bibinfo{person}{Nicholas Diakopoulos}.} \bibinfo{year}{2024}\natexlab{}.
\newblock \showarticletitle{Towards Leveraging News Media to Support Impact Assessment of AI Technologies}.
\newblock \bibinfo{journal}{\emph{arXiv preprint arXiv:2411.02536}} (\bibinfo{year}{2024}).
\newblock


\bibitem[Andriushchenko et~al\mbox{.}({[n.\,d.]})]%
        {andriushchenko273323256agentharm}
\bibfield{author}{\bibinfo{person}{Maksym Andriushchenko}, \bibinfo{person}{Alexandra Souly}, \bibinfo{person}{Mateusz Dziemian}, \bibinfo{person}{Derek Duenas}, \bibinfo{person}{Maxwell Lin}, \bibinfo{person}{Justin Wang}, \bibinfo{person}{Dan Hendrycks}, \bibinfo{person}{Andy Zou}, \bibinfo{person}{Zico Kolter}, \bibinfo{person}{Matt Fredrikson}, {et~al\mbox{.}}} \bibinfo{year}{[n.\,d.]}\natexlab{}.
\newblock \showarticletitle{Agentharm: A benchmark for measuring harmfulness of llm agents, 2024}.
\newblock \bibinfo{journal}{\emph{URL https://api. semanticscholar. org/CorpusID}}  \bibinfo{volume}{273323256} (\bibinfo{year}{[n.\,d.]}).
\newblock


\bibitem[Angwin et~al\mbox{.}(2022)]%
        {angwin_machine_2022}
\bibfield{author}{\bibinfo{person}{Julia Angwin}, \bibinfo{person}{Jeff Larson}, \bibinfo{person}{Surya Mattu}, {and} \bibinfo{person}{Lauren Kirchner}.} \bibinfo{year}{2022}\natexlab{}.
\newblock \showarticletitle{Machine bias}.
\newblock In \bibinfo{booktitle}{\emph{Ethics of data and analytics}}. \bibinfo{publisher}{Auerbach Publications}, \bibinfo{pages}{254--264}.
\newblock


\bibitem[{Anonymous}(2024)]%
        {anonymous2024}
\bibfield{author}{\bibinfo{person}{{Anonymous}}.} \bibinfo{year}{2024}\natexlab{}.
\newblock \bibinfo{title}{Enhancing LLMs for Governance with Human Oversight: Evaluating and Aligning LLMs on Expert Classification of Climate Misinformation for Detecting False or Misleading Claims about Climate Change}.
\newblock
\newblock
\newblock
\shownote{Under review}.


\bibitem[Avin et~al\mbox{.}(2020)]%
        {avin_exploring_2020}
\bibfield{author}{\bibinfo{person}{Shahar Avin}, \bibinfo{person}{Ross Gruetzemacher}, {and} \bibinfo{person}{James Fox}.} \bibinfo{year}{2020}\natexlab{}.
\newblock \showarticletitle{Exploring {AI} {Futures} {Through} {Role} {Play}}. In \bibinfo{booktitle}{\emph{Proceedings of the {AAAI}/{ACM} {Conference} on {AI}, {Ethics}, and {Society}}}. \bibinfo{publisher}{ACM}, \bibinfo{address}{New York NY USA}, \bibinfo{pages}{8--14}.
\newblock
\showISBNx{978-1-4503-7110-0}
\urldef\tempurl%
\url{https://doi.org/10.1145/3375627.3375817}
\showDOI{\tempurl}


\bibitem[Barnett and Diakopoulos(2022)]%
        {barnett_crowdsourcing_2022}
\bibfield{author}{\bibinfo{person}{Julia Barnett} {and} \bibinfo{person}{Nicholas Diakopoulos}.} \bibinfo{year}{2022}\natexlab{}.
\newblock \showarticletitle{Crowdsourcing {Impacts}: {Exploring} the {Utility} of {Crowds} for {Anticipating} {Societal} {Impacts} of {Algorithmic} {Decision} {Making}}. In \bibinfo{booktitle}{\emph{Proceedings of the 2022 {AAAI}/{ACM} {Conference} on {AI}, {Ethics}, and {Society}}}. \bibinfo{publisher}{ACM}, \bibinfo{address}{Oxford United Kingdom}, \bibinfo{pages}{56--67}.
\newblock
\showISBNx{978-1-4503-9247-1}
\urldef\tempurl%
\url{https://doi.org/10.1145/3514094.3534145}
\showDOI{\tempurl}


\bibitem[Barnett et~al\mbox{.}(2024)]%
        {barnett_simulating_2024-1}
\bibfield{author}{\bibinfo{person}{Julia Barnett}, \bibinfo{person}{Kimon Kieslich}, {and} \bibinfo{person}{Nicholas Diakopoulos}.} \bibinfo{year}{2024}\natexlab{}.
\newblock \showarticletitle{Simulating {Policy} {Impacts}: {Developing} a {Generative} {Scenario} {Writing} {Method} to {Evaluate} the {Perceived} {Effects} of {Regulation}}. In \bibinfo{booktitle}{\emph{Proceedings of the {AAAI}/{ACM} {Conference} on {AI}, {Ethics}, and {Society}}}, Vol.~\bibinfo{volume}{7}. \bibinfo{pages}{82--93}.
\newblock


\bibitem[Becker(2001)]%
        {becker_social_2001}
\bibfield{author}{\bibinfo{person}{Henk~A. Becker}.} \bibinfo{year}{2001}\natexlab{}.
\newblock \showarticletitle{Social impact assessment}.
\newblock \bibinfo{journal}{\emph{European Journal of Operational Research}} \bibinfo{volume}{128}, \bibinfo{number}{2} (\bibinfo{date}{Jan.} \bibinfo{year}{2001}), \bibinfo{pages}{311--321}.
\newblock
\showISSN{03772217}
\urldef\tempurl%
\url{https://doi.org/10.1016/S0377-2217(00)00074-6}
\showDOI{\tempurl}


\bibitem[Becker et~al\mbox{.}(2024)]%
        {becker2024policies}
\bibfield{author}{\bibinfo{person}{Kim~Bj{\"o}rn Becker}, \bibinfo{person}{Felix~M Simon}, {and} \bibinfo{person}{Christopher Crum}.} \bibinfo{year}{2024}\natexlab{}.
\newblock \showarticletitle{Policies in parallel? A comparative study of journalistic AI policies in 52 global news organisations}.
\newblock \bibinfo{journal}{\emph{Digital Journalism}} (\bibinfo{year}{2024}), \bibinfo{pages}{1--21}.
\newblock


\bibitem[Biden(2023)]%
        {biden2023executive}
\bibfield{author}{\bibinfo{person}{Joseph~R Biden}.} \bibinfo{year}{2023}\natexlab{}.
\newblock \showarticletitle{Executive order on the safe, secure, and trustworthy development and use of artificial intelligence}.
\newblock  (\bibinfo{year}{2023}).
\newblock


\bibitem[Bird et~al\mbox{.}(2023)]%
        {bird_typology_2023}
\bibfield{author}{\bibinfo{person}{Charlotte Bird}, \bibinfo{person}{Eddie~L. Ungless}, {and} \bibinfo{person}{Atoosa Kasirzadeh}.} \bibinfo{year}{2023}\natexlab{}.
\newblock \bibinfo{title}{Typology of {Risks} of {Generative} {Text}-to-{Image} {Models}}.
\newblock
\newblock
\urldef\tempurl%
\url{http://arxiv.org/abs/2307.05543}
\showURL{%
\tempurl}
\newblock
\shownote{arXiv:2307.05543 [cs]}.


\bibitem[Bommasani et~al\mbox{.}(2021)]%
        {bommasani2021opportunities}
\bibfield{author}{\bibinfo{person}{Rishi Bommasani}, \bibinfo{person}{Drew~A Hudson}, \bibinfo{person}{Ehsan Adeli}, \bibinfo{person}{Russ Altman}, \bibinfo{person}{Simran Arora}, \bibinfo{person}{Sydney von Arx}, \bibinfo{person}{Michael~S Bernstein}, \bibinfo{person}{Jeannette Bohg}, \bibinfo{person}{Antoine Bosselut}, \bibinfo{person}{Emma Brunskill}, {et~al\mbox{.}}} \bibinfo{year}{2021}\natexlab{}.
\newblock \showarticletitle{On the opportunities and risks of foundation models}.
\newblock \bibinfo{journal}{\emph{arXiv preprint arXiv:2108.07258}} (\bibinfo{year}{2021}).
\newblock


\bibitem[Bonaccorsi et~al\mbox{.}(2020)]%
        {bonaccorsi2020expert}
\bibfield{author}{\bibinfo{person}{Andrea Bonaccorsi}, \bibinfo{person}{Riccardo Apreda}, {and} \bibinfo{person}{Gualtiero Fantoni}.} \bibinfo{year}{2020}\natexlab{}.
\newblock \showarticletitle{Expert biases in technology foresight. Why they are a problem and how to mitigate them}.
\newblock \bibinfo{journal}{\emph{Technological Forecasting and Social Change}}  \bibinfo{volume}{151} (\bibinfo{year}{2020}), \bibinfo{pages}{119855}.
\newblock


\bibitem[Braun and Clarke(2012)]%
        {braun2012thematic}
\bibfield{author}{\bibinfo{person}{Virginia Braun} {and} \bibinfo{person}{Victoria Clarke}.} \bibinfo{year}{2012}\natexlab{}.
\newblock \bibinfo{booktitle}{\emph{Thematic analysis.}}
\newblock \bibinfo{publisher}{American Psychological Association}.
\newblock


\bibitem[Brennen(2018)]%
        {brennen_industry-led_2018}
\bibfield{author}{\bibinfo{person}{J. Brennen}.} \bibinfo{year}{2018}\natexlab{}.
\newblock \showarticletitle{An industry-led debate: {How} {UK} media cover artificial intelligence}.
\newblock  (\bibinfo{year}{2018}).
\newblock
\newblock
\shownote{Publisher: Reuters Institute for the Study of Journalism}.


\bibitem[Brennen et~al\mbox{.}(2021)]%
        {brennen2021balancing}
\bibfield{author}{\bibinfo{person}{J~Scott Brennen}, \bibinfo{person}{Philip~N Howard}, {and} \bibinfo{person}{Rasmus~K Nielsen}.} \bibinfo{year}{2021}\natexlab{}.
\newblock \showarticletitle{Balancing product reviews, traffic targets, and industry criticism: UK technology journalism in practice}.
\newblock \bibinfo{journal}{\emph{Journalism Practice}} \bibinfo{volume}{15}, \bibinfo{number}{10} (\bibinfo{year}{2021}), \bibinfo{pages}{1479--1496}.
\newblock


\bibitem[Brennen et~al\mbox{.}(2022)]%
        {brennen_what_2022}
\bibfield{author}{\bibinfo{person}{J~Scott Brennen}, \bibinfo{person}{Philip~N Howard}, {and} \bibinfo{person}{Rasmus~K Nielsen}.} \bibinfo{year}{2022}\natexlab{}.
\newblock \showarticletitle{What to expect when you’re expecting robots: {Futures}, expectations, and pseudo-artificial general intelligence in {UK} news}.
\newblock \bibinfo{journal}{\emph{Journalism}} \bibinfo{volume}{23}, \bibinfo{number}{1} (\bibinfo{date}{Jan.} \bibinfo{year}{2022}), \bibinfo{pages}{22--38}.
\newblock
\showISSN{1464-8849, 1741-3001}
\urldef\tempurl%
\url{https://doi.org/10.1177/1464884920947535}
\showDOI{\tempurl}


\bibitem[Bruckman(2020)]%
        {bruckman2020have}
\bibfield{author}{\bibinfo{person}{Amy Bruckman}.} \bibinfo{year}{2020}\natexlab{}.
\newblock \showarticletitle{'Have you thought about…' talking about ethical implications of research}.
\newblock \bibinfo{journal}{\emph{Commun. ACM}} \bibinfo{volume}{63}, \bibinfo{number}{9} (\bibinfo{year}{2020}), \bibinfo{pages}{38--40}.
\newblock


\bibitem[Bunz and Braghieri(2022)]%
        {bunz_ai_2022}
\bibfield{author}{\bibinfo{person}{Mercedes Bunz} {and} \bibinfo{person}{Marco Braghieri}.} \bibinfo{year}{2022}\natexlab{}.
\newblock \showarticletitle{The {AI} doctor will see you now: assessing the framing of {AI} in news coverage}.
\newblock \bibinfo{journal}{\emph{AI \& SOCIETY}} \bibinfo{volume}{37}, \bibinfo{number}{1} (\bibinfo{date}{March} \bibinfo{year}{2022}), \bibinfo{pages}{9--22}.
\newblock
\showISSN{0951-5666, 1435-5655}
\urldef\tempurl%
\url{https://doi.org/10.1007/s00146-021-01145-9}
\showDOI{\tempurl}


\bibitem[Burtell and Woodside(2023)]%
        {burtell2023artificial}
\bibfield{author}{\bibinfo{person}{Matthew Burtell} {and} \bibinfo{person}{Thomas Woodside}.} \bibinfo{year}{2023}\natexlab{}.
\newblock \showarticletitle{Artificial influence: An analysis of AI-driven persuasion}.
\newblock \bibinfo{journal}{\emph{arXiv preprint arXiv:2303.08721}} (\bibinfo{year}{2023}).
\newblock


\bibitem[Buçinca et~al\mbox{.}(2023)]%
        {bucinca_aha_2023}
\bibfield{author}{\bibinfo{person}{Zana Buçinca}, \bibinfo{person}{Chau~Minh Pham}, \bibinfo{person}{Maurice Jakesch}, \bibinfo{person}{Marco~Tulio Ribeiro}, \bibinfo{person}{Alexandra Olteanu}, {and} \bibinfo{person}{Saleema Amershi}.} \bibinfo{year}{2023}\natexlab{}.
\newblock \bibinfo{title}{{AHA}!: {Facilitating} {AI} {Impact} {Assessment} by {Generating} {Examples} of {Harms}}.
\newblock
\newblock
\urldef\tempurl%
\url{http://arxiv.org/abs/2306.03280}
\showURL{%
\tempurl}
\newblock
\shownote{arXiv:2306.03280 [cs]}.


\bibitem[Cave et~al\mbox{.}(2018)]%
        {cave_portrayals_2018}
\bibfield{author}{\bibinfo{person}{Stephen Cave}, \bibinfo{person}{Claire Craig}, \bibinfo{person}{Kanta Dihal}, \bibinfo{person}{Sarah Dillon}, \bibinfo{person}{Jessica Montgomery}, \bibinfo{person}{Beth Singler}, {and} \bibinfo{person}{Lindsay Taylor}.} \bibinfo{year}{2018}\natexlab{}.
\newblock \bibinfo{booktitle}{\emph{Portrayals and perceptions of {AI} and why they matter}}.
\newblock \bibinfo{type}{{T}echnical {R}eport}. \bibinfo{institution}{Apollo - University of Cambridge Repository}.
\newblock
\urldef\tempurl%
\url{https://doi.org/10.17863/cam.34502}
\showDOI{\tempurl}


\bibitem[Cave and ÓhÉigeartaigh(2018)]%
        {cave_ai_2018}
\bibfield{author}{\bibinfo{person}{Stephen Cave} {and} \bibinfo{person}{Seán~S. ÓhÉigeartaigh}.} \bibinfo{year}{2018}\natexlab{}.
\newblock \showarticletitle{An {AI} {Race} for {Strategic} {Advantage}: {Rhetoric} and {Risks}}. In \bibinfo{booktitle}{\emph{Proceedings of the 2018 {AAAI}/{ACM} {Conference} on {AI}, {Ethics}, and {Society}}}. \bibinfo{publisher}{ACM}, \bibinfo{address}{New Orleans LA USA}, \bibinfo{pages}{36--40}.
\newblock
\showISBNx{978-1-4503-6012-8}
\urldef\tempurl%
\url{https://doi.org/10.1145/3278721.3278780}
\showDOI{\tempurl}


\bibitem[Check(2025)]%
        {mediabias2025}
\bibfield{author}{\bibinfo{person}{Media~Bias/Fact Check}.} \bibinfo{year}{2025}\natexlab{}.
\newblock \bibinfo{title}{Media Bias/Fact Check: The Most Comprehensive Media Bias Resource}.
\newblock
\newblock
\urldef\tempurl%
\url{https://mediabiasfactcheck.com/}
\showURL{%
\tempurl}
\newblock
\shownote{Accessed on January 17, 2025}.


\bibitem[Chinn et~al\mbox{.}(2020)]%
        {chinn2020politicization}
\bibfield{author}{\bibinfo{person}{Sedona Chinn}, \bibinfo{person}{P~Sol Hart}, {and} \bibinfo{person}{Stuart Soroka}.} \bibinfo{year}{2020}\natexlab{}.
\newblock \showarticletitle{Politicization and polarization in climate change news content, 1985-2017}.
\newblock \bibinfo{journal}{\emph{Science Communication}} \bibinfo{volume}{42}, \bibinfo{number}{1} (\bibinfo{year}{2020}), \bibinfo{pages}{112--129}.
\newblock


\bibitem[Chuan et~al\mbox{.}(2019)]%
        {chuan2019framing}
\bibfield{author}{\bibinfo{person}{Ching-Hua Chuan}, \bibinfo{person}{Wan-Hsiu~Sunny Tsai}, {and} \bibinfo{person}{Su~Yeon Cho}.} \bibinfo{year}{2019}\natexlab{}.
\newblock \showarticletitle{Framing artificial intelligence in American newspapers}. In \bibinfo{booktitle}{\emph{Proceedings of the 2019 AAAI/ACM Conference on AI, Ethics, and Society}}. \bibinfo{pages}{339--344}.
\newblock


\bibitem[Cohen and Waldman(2023)]%
        {cohen_introduction_2023}
\bibfield{author}{\bibinfo{person}{Julie~E Cohen} {and} \bibinfo{person}{Ari~Azra Waldman}.} \bibinfo{year}{2023}\natexlab{}.
\newblock \showarticletitle{Introduction: {Framing} {Regulatory} {Managerialism} as an {Object} of {Study} and {Strategic} {Displacement}}.
\newblock \bibinfo{journal}{\emph{Law \& Contemp. Probs.}}  \bibinfo{volume}{86} (\bibinfo{year}{2023}), \bibinfo{pages}{i}.
\newblock
\newblock
\shownote{Publisher: HeinOnline}.


\bibitem[Constantaras et~al\mbox{.}(2023)]%
        {constantaras_inside_2023}
\bibfield{author}{\bibinfo{person}{Eva Constantaras}, \bibinfo{person}{Gabriel Geiger}, \bibinfo{person}{Justin-Casimir Braun}, \bibinfo{person}{Dhruv Mehrotra}, {and} \bibinfo{person}{Hiet Aung}.} \bibinfo{year}{2023}\natexlab{}.
\newblock \bibinfo{booktitle}{\emph{Inside the suspicion machine.}}
\newblock \bibinfo{type}{{T}echnical {R}eport}.
\newblock
\urldef\tempurl%
\url{https://www.wired.com/story/welfare-state-algorithms/}
\showURL{%
\tempurl}


\bibitem[Costanza-Chock et~al\mbox{.}(2022)]%
        {costanza-chock_who_2022}
\bibfield{author}{\bibinfo{person}{Sasha Costanza-Chock}, \bibinfo{person}{Inioluwa~Deborah Raji}, {and} \bibinfo{person}{Joy Buolamwini}.} \bibinfo{year}{2022}\natexlab{}.
\newblock \showarticletitle{Who {Audits} the {Auditors}? {Recommendations} from a field scan of the algorithmic auditing ecosystem}. In \bibinfo{booktitle}{\emph{2022 {ACM} {Conference} on {Fairness}, {Accountability}, and {Transparency}}}. \bibinfo{publisher}{ACM}, \bibinfo{address}{Seoul Republic of Korea}, \bibinfo{pages}{1571--1583}.
\newblock
\showISBNx{978-1-4503-9352-2}
\urldef\tempurl%
\url{https://doi.org/10.1145/3531146.3533213}
\showDOI{\tempurl}


\bibitem[Crawford(2016)]%
        {crawford2016artificial}
\bibfield{author}{\bibinfo{person}{Kate Crawford}.} \bibinfo{year}{2016}\natexlab{}.
\newblock \showarticletitle{Artificial intelligence’s white guy problem}.
\newblock \bibinfo{journal}{\emph{The New York Times}} \bibinfo{volume}{25}, \bibinfo{number}{06} (\bibinfo{year}{2016}), \bibinfo{pages}{5}.
\newblock


\bibitem[Crawford(2021)]%
        {crawford_atlas_2021}
\bibfield{author}{\bibinfo{person}{Kate Crawford}.} \bibinfo{year}{2021}\natexlab{}.
\newblock \bibinfo{booktitle}{\emph{The atlas of {AI}: {Power}, politics, and the planetary costs of artificial intelligence}}.
\newblock \bibinfo{publisher}{Yale University Press}.
\newblock


\bibitem[De~Angelis et~al\mbox{.}(2023)]%
        {de2023chatgpt}
\bibfield{author}{\bibinfo{person}{Luigi De~Angelis}, \bibinfo{person}{Francesco Baglivo}, \bibinfo{person}{Guglielmo Arzilli}, \bibinfo{person}{Gaetano~Pierpaolo Privitera}, \bibinfo{person}{Paolo Ferragina}, \bibinfo{person}{Alberto~Eugenio Tozzi}, {and} \bibinfo{person}{Caterina Rizzo}.} \bibinfo{year}{2023}\natexlab{}.
\newblock \showarticletitle{ChatGPT and the rise of large language models: the new AI-driven infodemic threat in public health}.
\newblock \bibinfo{journal}{\emph{Frontiers in public health}}  \bibinfo{volume}{11} (\bibinfo{year}{2023}), \bibinfo{pages}{1166120}.
\newblock


\bibitem[De~Wilde(2011)]%
        {de2011no}
\bibfield{author}{\bibinfo{person}{Pieter De~Wilde}.} \bibinfo{year}{2011}\natexlab{}.
\newblock \showarticletitle{No polity for old politics? A framework for analyzing the politicization of European integration}.
\newblock \bibinfo{journal}{\emph{Journal of European integration}} \bibinfo{volume}{33}, \bibinfo{number}{5} (\bibinfo{year}{2011}), \bibinfo{pages}{559--575}.
\newblock


\bibitem[Diakopoulos(2015)]%
        {diakopoulos_2015}
\bibfield{author}{\bibinfo{person}{Nicholas Diakopoulos}.} \bibinfo{year}{2015}\natexlab{}.
\newblock \showarticletitle{{Algorithmic Accountability: Journalistic investigation of computational power structures}}.
\newblock \bibinfo{journal}{\emph{Digital Journalism}} \bibinfo{volume}{3}, \bibinfo{number}{3} (\bibinfo{year}{2015}), \bibinfo{pages}{398 -- 415}.
\newblock
\urldef\tempurl%
\url{https://doi.org/10.1080/21670811.2014.976411}
\showDOI{\tempurl}


\bibitem[Diakopoulos(2025)]%
        {diakopoulos_2025}
\bibfield{author}{\bibinfo{person}{Nicholas Diakopoulos}.} \bibinfo{year}{2025}\natexlab{}.
\newblock \showarticletitle{{Prospective Algorithmic Accountability and the Role of the News Media}}.
\newblock In \bibinfo{booktitle}{\emph{Computer Ethics Across Disciplines: Deborah G. Johnson and Algorithmic Accountability}}, \bibfield{editor}{\bibinfo{person}{M.~Moorman} {and} \bibinfo{person}{M.~Verdicchio}} (Eds.).
\newblock


\bibitem[Do et~al\mbox{.}(2023)]%
        {do2023s}
\bibfield{author}{\bibinfo{person}{Kimberly Do}, \bibinfo{person}{Rock~Yuren Pang}, \bibinfo{person}{Jiachen Jiang}, {and} \bibinfo{person}{Katharina Reinecke}.} \bibinfo{year}{2023}\natexlab{}.
\newblock \showarticletitle{“That’s important, but...”: How Computer Science Researchers Anticipate Unintended Consequences of Their Research Innovations}. In \bibinfo{booktitle}{\emph{Proceedings of the 2023 CHI Conference on Human Factors in Computing Systems}}. \bibinfo{pages}{1--16}.
\newblock


\bibitem[Documentation(2025)]%
        {hdbscan2025}
\bibfield{author}{\bibinfo{person}{HDBSCAN Documentation}.} \bibinfo{year}{2025}\natexlab{}.
\newblock \bibinfo{title}{HDBSCAN: Hierarchical Density-Based Spatial Clustering}.
\newblock
\newblock
\urldef\tempurl%
\url{https://hdbscan.readthedocs.io/}
\showURL{%
\tempurl}
\newblock
\shownote{Accessed on January 17, 2025}.


\bibitem[Dotan et~al\mbox{.}(2024)]%
        {dotan_evolving_2024}
\bibfield{author}{\bibinfo{person}{Ravit Dotan}, \bibinfo{person}{Borhane Blili-Hamelin}, \bibinfo{person}{Ravi Madhavan}, \bibinfo{person}{Jeanna Matthews}, {and} \bibinfo{person}{Joshua Scarpino}.} \bibinfo{year}{2024}\natexlab{}.
\newblock \bibinfo{title}{Evolving {AI} {Risk} {Management}: {A} {Maturity} {Model} based on the {NIST} {AI} {Risk} {Management} {Framework}}.
\newblock
\newblock
\urldef\tempurl%
\url{http://arxiv.org/abs/2401.15229}
\showURL{%
\tempurl}
\newblock
\shownote{arXiv:2401.15229 [cs]}.


\bibitem[Fast and Horvitz(2017)]%
        {fast_long-term_2017}
\bibfield{author}{\bibinfo{person}{Ethan Fast} {and} \bibinfo{person}{Eric Horvitz}.} \bibinfo{year}{2017}\natexlab{}.
\newblock \showarticletitle{Long-term trends in the public perception of artificial intelligence}. In \bibinfo{booktitle}{\emph{Proceedings of the {AAAI} conference on artificial intelligence}}, Vol.~\bibinfo{volume}{31}.
\newblock
\newblock
\shownote{Issue: 1}.


\bibitem[Filippucci et~al\mbox{.}(2024)]%
        {filippucci2024impact}
\bibfield{author}{\bibinfo{person}{Francesco Filippucci}, \bibinfo{person}{Peter Gal}, \bibinfo{person}{Cecilia Jona-Lasinio}, \bibinfo{person}{Alvaro Leandro}, {and} \bibinfo{person}{Giuseppe Nicoletti}.} \bibinfo{year}{2024}\natexlab{}.
\newblock \showarticletitle{The impact of Artificial Intelligence on productivity, distribution and growth: Key mechanisms, initial evidence and policy challenges}.
\newblock  (\bibinfo{year}{2024}).
\newblock


\bibitem[Floridi and Strait(2020)]%
        {floridi_ethical_2020}
\bibfield{author}{\bibinfo{person}{Luciano Floridi} {and} \bibinfo{person}{Andrew Strait}.} \bibinfo{year}{2020}\natexlab{}.
\newblock \showarticletitle{Ethical {Foresight} {Analysis}: {What} it is and {Why} it is {Needed}?}
\newblock \bibinfo{journal}{\emph{Minds and Machines}} \bibinfo{volume}{30}, \bibinfo{number}{1} (\bibinfo{date}{March} \bibinfo{year}{2020}), \bibinfo{pages}{77--97}.
\newblock
\showISSN{0924-6495, 1572-8641}
\urldef\tempurl%
\url{https://doi.org/10.1007/s11023-020-09521-y}
\showDOI{\tempurl}


\bibitem[Gilardi et~al\mbox{.}(2024)]%
        {gilardi_we_2024}
\bibfield{author}{\bibinfo{person}{Fabrizio Gilardi}, \bibinfo{person}{Atoosa Kasirzadeh}, \bibinfo{person}{Abraham Bernstein}, \bibinfo{person}{Steffen Staab}, {and} \bibinfo{person}{Anita Gohdes}.} \bibinfo{year}{2024}\natexlab{}.
\newblock \showarticletitle{We need to understand the effect of narratives about generative {AI}}.
\newblock \bibinfo{journal}{\emph{Nature Human Behaviour}} (\bibinfo{date}{Oct.} \bibinfo{year}{2024}).
\newblock
\showISSN{2397-3374}
\urldef\tempurl%
\url{https://doi.org/10.1038/s41562-024-02026-z}
\showDOI{\tempurl}


\bibitem[Griffin(2024)]%
        {griffin_what_2024}
\bibfield{author}{\bibinfo{person}{Rachel Griffin}.} \bibinfo{year}{2024}\natexlab{}.
\newblock \bibinfo{title}{What do we talk about when we talk about risk? {Risk} politics in the {EU}’s {Digital} {Services} {Act}}.
\newblock
\newblock
\urldef\tempurl%
\url{https://dsa-observatory.eu/2024/07/31/what-do-we-talk-about-when-we-talk-about-risk-risk-politics-in-the-eus-digital-services-act/}
\showURL{%
\tempurl}


\bibitem[Hagerty and Rubinov(2019)]%
        {hagerty2019global}
\bibfield{author}{\bibinfo{person}{Alexa Hagerty} {and} \bibinfo{person}{Igor Rubinov}.} \bibinfo{year}{2019}\natexlab{}.
\newblock \showarticletitle{Global AI ethics: a review of the social impacts and ethical implications of artificial intelligence}.
\newblock \bibinfo{journal}{\emph{arXiv preprint arXiv:1907.07892}} (\bibinfo{year}{2019}).
\newblock


\bibitem[Hartmann et~al\mbox{.}(2024)]%
        {hartmann_addressing_2024}
\bibfield{author}{\bibinfo{person}{David Hartmann}, \bibinfo{person}{José Renato~Laranjeira De~Pereira}, \bibinfo{person}{Chiara Streitbörger}, {and} \bibinfo{person}{Bettina Berendt}.} \bibinfo{year}{2024}\natexlab{}.
\newblock \showarticletitle{Addressing the regulatory gap: moving towards an {EU} {AI} audit ecosystem beyond the {AI} {Act} by including civil society}.
\newblock \bibinfo{journal}{\emph{AI and Ethics}} (\bibinfo{date}{Nov.} \bibinfo{year}{2024}).
\newblock
\showISSN{2730-5953, 2730-5961}
\urldef\tempurl%
\url{https://doi.org/10.1007/s43681-024-00595-3}
\showDOI{\tempurl}


\bibitem[Hautala and Heino(2023)]%
        {hautala_spectrum_2023}
\bibfield{author}{\bibinfo{person}{Johanna Hautala} {and} \bibinfo{person}{Hanna Heino}.} \bibinfo{year}{2023}\natexlab{}.
\newblock \showarticletitle{Spectrum of {AI} futures imaginaries by {AI} practitioners in {Finland} and {Singapore}: {The} unimagined speed of {AI} progress}.
\newblock \bibinfo{journal}{\emph{Futures}}  \bibinfo{volume}{153} (\bibinfo{date}{Oct.} \bibinfo{year}{2023}), \bibinfo{pages}{103247}.
\newblock
\showISSN{00163287}
\urldef\tempurl%
\url{https://doi.org/10.1016/j.futures.2023.103247}
\showDOI{\tempurl}


\bibitem[Herdel et~al\mbox{.}(2024)]%
        {herdel2024exploregen}
\bibfield{author}{\bibinfo{person}{Viviane Herdel}, \bibinfo{person}{Sanja {\v{S}}{\'c}epanovi{\'c}}, \bibinfo{person}{Edyta Bogucka}, {and} \bibinfo{person}{Daniele Quercia}.} \bibinfo{year}{2024}\natexlab{}.
\newblock \showarticletitle{ExploreGen: Large language models for envisioning the uses and risks of AI technologies}. In \bibinfo{booktitle}{\emph{Proceedings of the AAAI/ACM Conference on AI, Ethics, and Society}}, Vol.~\bibinfo{volume}{7}. \bibinfo{pages}{584--596}.
\newblock


\bibitem[Hernandes and Corsi(2024)]%
        {hernandes2024auditing}
\bibfield{author}{\bibinfo{person}{Raphael Hernandes} {and} \bibinfo{person}{Giulio Corsi}.} \bibinfo{year}{2024}\natexlab{}.
\newblock \showarticletitle{Auditing Google's Search Algorithm: Measuring News Diversity Across Brazil, the UK, and the US}.
\newblock \bibinfo{journal}{\emph{arXiv preprint arXiv:2410.23842}} (\bibinfo{year}{2024}).
\newblock


\bibitem[Hilgard and Li(2017)]%
        {jamieson_recap_2017}
\bibfield{author}{\bibinfo{person}{Joseph Hilgard} {and} \bibinfo{person}{Nan Li}.} \bibinfo{year}{2017}\natexlab{}.
\newblock \bibinfo{booktitle}{\emph{A {Recap}}}. Vol.~\bibinfo{volume}{1}.
\newblock \bibinfo{publisher}{Oxford University Press}.
\newblock
\urldef\tempurl%
\url{https://doi.org/10.1093/oxfordhb/9780190497620.013.8}
\showDOI{\tempurl}


\bibitem[Huang et~al\mbox{.}(2024)]%
        {huang2024uncovering}
\bibfield{author}{\bibinfo{person}{Hong Huang}, \bibinfo{person}{Hua Zhu}, \bibinfo{person}{Wenshi Liu}, \bibinfo{person}{Hua Gao}, \bibinfo{person}{Hai Jin}, {and} \bibinfo{person}{Bang Liu}.} \bibinfo{year}{2024}\natexlab{}.
\newblock \showarticletitle{Uncovering the essence of diverse media biases from the semantic embedding space}.
\newblock \bibinfo{journal}{\emph{Humanities and Social Sciences Communications}} \bibinfo{volume}{11}, \bibinfo{number}{1} (\bibinfo{year}{2024}), \bibinfo{pages}{1--12}.
\newblock


\bibitem[Ittefaq et~al\mbox{.}(2025)]%
        {ittefaq_global_2025}
\bibfield{author}{\bibinfo{person}{Muhammad Ittefaq}, \bibinfo{person}{Ali Zain}, \bibinfo{person}{Rauf Arif}, \bibinfo{person}{Mohammad Ala-Uddin}, \bibinfo{person}{Taufiq Ahmad}, {and} \bibinfo{person}{Azhar Iqbal}.} \bibinfo{year}{2025}\natexlab{}.
\newblock \showarticletitle{Global news media coverage of artificial intelligence ({AI}): {A} comparative analysis of frames, sentiments, and trends across 12 countries}.
\newblock \bibinfo{journal}{\emph{Telematics and Informatics}}  \bibinfo{volume}{96} (\bibinfo{date}{Jan.} \bibinfo{year}{2025}), \bibinfo{pages}{102223}.
\newblock
\showISSN{07365853}
\urldef\tempurl%
\url{https://doi.org/10.1016/j.tele.2024.102223}
\showDOI{\tempurl}


\bibitem[Jobin et~al\mbox{.}(2019)]%
        {jobin_global_2019}
\bibfield{author}{\bibinfo{person}{Anna Jobin}, \bibinfo{person}{Marcello Ienca}, {and} \bibinfo{person}{Effy Vayena}.} \bibinfo{year}{2019}\natexlab{}.
\newblock \showarticletitle{The global landscape of {AI} ethics guidelines}.
\newblock \bibinfo{journal}{\emph{Nature Machine Intelligence}} \bibinfo{volume}{1}, \bibinfo{number}{9} (\bibinfo{date}{Sept.} \bibinfo{year}{2019}), \bibinfo{pages}{389--399}.
\newblock
\showISSN{2522-5839}
\urldef\tempurl%
\url{https://doi.org/10.1038/s42256-019-0088-2}
\showDOI{\tempurl}


\bibitem[Kieslich(2024)]%
        {kieslich_role_2024}
\bibfield{author}{\bibinfo{person}{Kimon Kieslich}.} \bibinfo{year}{2024}\natexlab{}.
\newblock \emph{\bibinfo{title}{The role of public opinion on ethical {AI} principles and its implication for a common good-oriented implementation.}}
\newblock \bibinfo{thesistype}{Ph.\,D. Dissertation}. \bibinfo{school}{Universität Hohenheim}.
\newblock
\urldef\tempurl%
\url{https://hohpublica.uni-hohenheim.de/handle/123456789/15939}
\showURL{%
\tempurl}


\bibitem[Kieslich et~al\mbox{.}(2024a)]%
        {kieslich_anticipating_2024}
\bibfield{author}{\bibinfo{person}{Kimon Kieslich}, \bibinfo{person}{Nicholas Diakopoulos}, {and} \bibinfo{person}{Natali Helberger}.} \bibinfo{year}{2024}\natexlab{a}.
\newblock \showarticletitle{Anticipating impacts: using large-scale scenario-writing to explore diverse implications of generative {AI} in the news environment}.
\newblock \bibinfo{journal}{\emph{AI and Ethics}} (\bibinfo{date}{May} \bibinfo{year}{2024}).
\newblock
\showISSN{2730-5953, 2730-5961}
\urldef\tempurl%
\url{https://doi.org/10.1007/s43681-024-00497-4}
\showDOI{\tempurl}


\bibitem[Kieslich et~al\mbox{.}(2024b)]%
        {kieslich_using_2024}
\bibfield{author}{\bibinfo{person}{Kimon Kieslich}, \bibinfo{person}{Nicholas Diakopoulos}, {and} \bibinfo{person}{Natali Helberger}.} \bibinfo{year}{2024}\natexlab{b}.
\newblock \bibinfo{title}{Using {Scenario}-{Writing} for {Identifying} and {Mitigating} {Impacts} of {Generative} {AI}}.
\newblock
\newblock
\urldef\tempurl%
\url{https://doi.org/10.48550/ARXIV.2410.23704}
\showDOI{\tempurl}
\newblock
\shownote{Version Number: 1}.


\bibitem[Kieslich et~al\mbox{.}(2022)]%
        {kieslich_everything_2022}
\bibfield{author}{\bibinfo{person}{Kimon Kieslich}, \bibinfo{person}{Pero Došenović}, {and} \bibinfo{person}{Frank Marcinkowski}.} \bibinfo{year}{2022}\natexlab{}.
\newblock \bibinfo{booktitle}{\emph{Everything, but hardly any science fiction}}.
\newblock \bibinfo{type}{{T}echnical {R}eport}~7. \bibinfo{institution}{Meinungsmonitor Künstliche Intelligenz}.
\newblock
\urldef\tempurl%
\url{https://www.researchgate.net/profile/Kimon-Kieslich/publication/365033703_Everything_but_hardly_any_science_fiction/links/63638442431b1f5300685b2d/Everything-but-hardly-any-science-fiction.pdf}
\showURL{%
\tempurl}


\bibitem[Kieslich et~al\mbox{.}(2024c)]%
        {kieslich_my_2024}
\bibfield{author}{\bibinfo{person}{Kimon Kieslich}, \bibinfo{person}{Natali Helberger}, {and} \bibinfo{person}{Nicholas Diakopoulos}.} \bibinfo{year}{2024}\natexlab{c}.
\newblock \showarticletitle{My {Future} with {My} {Chatbot}: {A} {Scenario}-{Driven}, {User}-{Centric} {Approach} to {Anticipating} {AI} {Impacts}}. In \bibinfo{booktitle}{\emph{The 2024 {ACM} {Conference} on {Fairness}, {Accountability}, and {Transparency}}}. \bibinfo{publisher}{ACM}, \bibinfo{address}{Rio de Janeiro Brazil}, \bibinfo{pages}{2071--2085}.
\newblock
\showISBNx{9798400704505}
\urldef\tempurl%
\url{https://doi.org/10.1145/3630106.3659026}
\showDOI{\tempurl}


\bibitem[Kieslich et~al\mbox{.}(2023)]%
        {kieslich_ever_2023}
\bibfield{author}{\bibinfo{person}{Kimon Kieslich}, \bibinfo{person}{Marco Lünich}, {and} \bibinfo{person}{Pero Došenović}.} \bibinfo{year}{2023}\natexlab{}.
\newblock \showarticletitle{Ever {Heard} of {Ethical} {AI}? {Investigating} the {Salience} of {Ethical} {AI} {Issues} among the {German} {Population}}.
\newblock \bibinfo{journal}{\emph{International Journal of Human–Computer Interaction}} (\bibinfo{date}{Feb.} \bibinfo{year}{2023}), \bibinfo{pages}{1--14}.
\newblock
\showISSN{1044-7318, 1532-7590}
\urldef\tempurl%
\url{https://doi.org/10.1080/10447318.2023.2178612}
\showDOI{\tempurl}


\bibitem[Krafft et~al\mbox{.}(2022)]%
        {krafft_how_2022}
\bibfield{author}{\bibinfo{person}{Tobias~D. Krafft}, \bibinfo{person}{Katharina~A. Zweig}, {and} \bibinfo{person}{Pascal~D. König}.} \bibinfo{year}{2022}\natexlab{}.
\newblock \showarticletitle{How to regulate algorithmic decision‐making: {A} framework of regulatory requirements for different applications}.
\newblock \bibinfo{journal}{\emph{Regulation \& Governance}} \bibinfo{volume}{16}, \bibinfo{number}{1} (\bibinfo{date}{Jan.} \bibinfo{year}{2022}), \bibinfo{pages}{119--136}.
\newblock
\showISSN{1748-5983, 1748-5991}
\urldef\tempurl%
\url{https://doi.org/10.1111/rego.12369}
\showDOI{\tempurl}


\bibitem[Leetaru and Schrodt(2013)]%
        {leetaru2013gdelt}
\bibfield{author}{\bibinfo{person}{Kalev Leetaru} {and} \bibinfo{person}{Philip~A Schrodt}.} \bibinfo{year}{2013}\natexlab{}.
\newblock \showarticletitle{Gdelt: Global data on events, location, and tone, 1979--2012}. In \bibinfo{booktitle}{\emph{ISA annual convention}}, Vol.~\bibinfo{volume}{2}. Citeseer, \bibinfo{pages}{1--49}.
\newblock


\bibitem[Lin et~al\mbox{.}(2023)]%
        {lin2023high}
\bibfield{author}{\bibinfo{person}{Hause Lin}, \bibinfo{person}{Jana Lasser}, \bibinfo{person}{Stephan Lewandowsky}, \bibinfo{person}{Rocky Cole}, \bibinfo{person}{Andrew Gully}, \bibinfo{person}{David~G Rand}, {and} \bibinfo{person}{Gordon Pennycook}.} \bibinfo{year}{2023}\natexlab{}.
\newblock \showarticletitle{High level of correspondence across different news domain quality rating sets}.
\newblock \bibinfo{journal}{\emph{PNAS nexus}} \bibinfo{volume}{2}, \bibinfo{number}{9} (\bibinfo{year}{2023}), \bibinfo{pages}{pgad286}.
\newblock


\bibitem[Madiega(2021)]%
        {madiega2021artificial}
\bibfield{author}{\bibinfo{person}{Tambiama Madiega}.} \bibinfo{year}{2021}\natexlab{}.
\newblock \showarticletitle{Artificial intelligence act}.
\newblock \bibinfo{journal}{\emph{European Parliament: European Parliamentary Research Service}} (\bibinfo{year}{2021}).
\newblock


\bibitem[Mamasoliev(2024)]%
        {mamasoliev2024impact}
\bibfield{author}{\bibinfo{person}{Saidbek Mamasoliev}.} \bibinfo{year}{2024}\natexlab{}.
\newblock \showarticletitle{IMPACT OF ARTIFICIAL INTELLIGENCE ON US ECONOMIC GROWTH AND GLOBAL COMPETITIVENESS}.
\newblock \bibinfo{journal}{\emph{AMERICAN JOURNAL OF BUSINESS MANAGEMENT}} \bibinfo{volume}{2}, \bibinfo{number}{3} (\bibinfo{year}{2024}), \bibinfo{pages}{82--91}.
\newblock


\bibitem[Marzi et~al\mbox{.}(2024)]%
        {marzi2024k}
\bibfield{author}{\bibinfo{person}{Giacomo Marzi}, \bibinfo{person}{Marco Balzano}, {and} \bibinfo{person}{Davide Marchiori}.} \bibinfo{year}{2024}\natexlab{}.
\newblock \showarticletitle{K-Alpha Calculator--Krippendorff's Alpha Calculator: A user-friendly tool for computing Krippendorff's Alpha inter-rater reliability coefficient}.
\newblock \bibinfo{journal}{\emph{MethodsX}}  \bibinfo{volume}{12} (\bibinfo{year}{2024}), \bibinfo{pages}{102545}.
\newblock


\bibitem[McCombs and Shaw(1972)]%
        {mccombs1972agenda}
\bibfield{author}{\bibinfo{person}{Maxwell~E McCombs} {and} \bibinfo{person}{Donald~L Shaw}.} \bibinfo{year}{1972}\natexlab{}.
\newblock \showarticletitle{The agenda-setting function of mass media}.
\newblock \bibinfo{journal}{\emph{Public opinion quarterly}} \bibinfo{volume}{36}, \bibinfo{number}{2} (\bibinfo{year}{1972}), \bibinfo{pages}{176--187}.
\newblock


\bibitem[McGregor(2021)]%
        {mcgregor2021preventing}
\bibfield{author}{\bibinfo{person}{Sean McGregor}.} \bibinfo{year}{2021}\natexlab{}.
\newblock \showarticletitle{Preventing repeated real world AI failures by cataloging incidents: The AI incident database}. In \bibinfo{booktitle}{\emph{Proceedings of the AAAI Conference on Artificial Intelligence}}, Vol.~\bibinfo{volume}{35}. \bibinfo{pages}{15458--15463}.
\newblock


\bibitem[McInnes et~al\mbox{.}(2018)]%
        {mcinnes2018umap}
\bibfield{author}{\bibinfo{person}{Leland McInnes}, \bibinfo{person}{John Healy}, {and} \bibinfo{person}{James Melville}.} \bibinfo{year}{2018}\natexlab{}.
\newblock \showarticletitle{Umap: Uniform manifold approximation and projection for dimension reduction}.
\newblock \bibinfo{journal}{\emph{arXiv preprint arXiv:1802.03426}} (\bibinfo{year}{2018}).
\newblock


\bibitem[{Media Cloud}(2025)]%
        {mediacloud2025}
\bibfield{author}{\bibinfo{person}{{Media Cloud}}.} \bibinfo{year}{2025}\natexlab{}.
\newblock \bibinfo{title}{Collection 9272347}.
\newblock
\newblock
\urldef\tempurl%
\url{https://search.mediacloud.org/collections/9272347}
\showURL{%
\tempurl}
\newblock
\shownote{Accessed on January 17, 2025}.


\bibitem[Meißner(2024)]%
        {meisner_risks_2024}
\bibfield{author}{\bibinfo{person}{Florian Meißner}.} \bibinfo{year}{2024}\natexlab{}.
\newblock \showarticletitle{Risks and opportunities of ‘generative {A}.{I}.’: {How} do news media cover {ChatGPT}?}. In \bibinfo{booktitle}{\emph{International {Crisis} and {Risk} {Communication} {Conference} {Proceedings}}}. \bibinfo{publisher}{International Crisis and Risk Communication Association}.
\newblock
\urldef\tempurl%
\url{https://doi.org/10.69931/WFEY9011}
\showDOI{\tempurl}


\bibitem[Metcalf et~al\mbox{.}(2021)]%
        {metcalf_algorithmic_2021}
\bibfield{author}{\bibinfo{person}{Jacob Metcalf}, \bibinfo{person}{Emanuel Moss}, \bibinfo{person}{Elizabeth~Anne Watkins}, \bibinfo{person}{Ranjit Singh}, {and} \bibinfo{person}{Madeleine~Clare Elish}.} \bibinfo{year}{2021}\natexlab{}.
\newblock \showarticletitle{Algorithmic impact assessments and accountability: the co-construction of impacts}.
\newblock \bibinfo{journal}{\emph{Proceedings of the 2021 ACM Conference on Fairness, Accountability, and Transparency}} (\bibinfo{year}{2021}).
\newblock
\urldef\tempurl%
\url{https://doi.org/10.1145/3442188.3445935}
\showDOI{\tempurl}


\bibitem[Moss et~al\mbox{.}(2021)]%
        {moss_assembling_2021}
\bibfield{author}{\bibinfo{person}{Emanuel Moss}, \bibinfo{person}{Elizabeth Watkins}, \bibinfo{person}{Ranjit Singh}, \bibinfo{person}{Madeleine~Clare Elish}, {and} \bibinfo{person}{Jacob Metcalf}.} \bibinfo{year}{2021}\natexlab{}.
\newblock \showarticletitle{Assembling {Accountability}: {Algorithmic} {Impact} {Assessment} for the {Public} {Interest}}.
\newblock \bibinfo{journal}{\emph{SSRN Electronic Journal}} (\bibinfo{year}{2021}).
\newblock
\showISSN{1556-5068}
\urldef\tempurl%
\url{https://doi.org/10.2139/ssrn.3877437}
\showDOI{\tempurl}


\bibitem[Nanayakkara et~al\mbox{.}(2021)]%
        {nanayakkara_unpacking_2021}
\bibfield{author}{\bibinfo{person}{Priyanka Nanayakkara}, \bibinfo{person}{Jessica Hullman}, {and} \bibinfo{person}{Nicholas Diakopoulos}.} \bibinfo{year}{2021}\natexlab{}.
\newblock \showarticletitle{Unpacking the expressed consequences of {AI} research in broader impact statements}. In \bibinfo{booktitle}{\emph{Proceedings of the 2021 {AAAI}/{ACM} {Conference} on {AI}, {Ethics}, and {Society}}}. \bibinfo{pages}{795--806}.
\newblock


\bibitem[Nechushtai et~al\mbox{.}(2024)]%
        {nechushtai2024more}
\bibfield{author}{\bibinfo{person}{Efrat Nechushtai}, \bibinfo{person}{Rodrigo Zamith}, {and} \bibinfo{person}{Seth~C Lewis}.} \bibinfo{year}{2024}\natexlab{}.
\newblock \showarticletitle{More of the same? Homogenization in news recommendations when users search on Google, YouTube, Facebook, and Twitter}.
\newblock \bibinfo{journal}{\emph{Mass Communication and Society}} \bibinfo{volume}{27}, \bibinfo{number}{6} (\bibinfo{year}{2024}), \bibinfo{pages}{1309--1335}.
\newblock


\bibitem[Nguyen and Hekman(2022)]%
        {nguyen_new_2022}
\bibfield{author}{\bibinfo{person}{Dennis Nguyen} {and} \bibinfo{person}{Erik Hekman}.} \bibinfo{year}{2022}\natexlab{}.
\newblock \showarticletitle{A ‘{New} {Arms} {Race}’? {Framing} {China} and the {U}.{S}.{A}. in {A}.{I}. {News} {Reporting}: {A} {Comparative} {Analysis} of the {Washington} {Post} and {South} {China} {Morning} {Post}}.
\newblock \bibinfo{journal}{\emph{Global Media and China}} \bibinfo{volume}{7}, \bibinfo{number}{1} (\bibinfo{date}{March} \bibinfo{year}{2022}), \bibinfo{pages}{58--77}.
\newblock
\showISSN{2059-4364, 2059-4372}
\urldef\tempurl%
\url{https://doi.org/10.1177/20594364221078626}
\showDOI{\tempurl}


\bibitem[Nguyen and Hekman(2024)]%
        {nguyen_news_2024}
\bibfield{author}{\bibinfo{person}{Dennis Nguyen} {and} \bibinfo{person}{Erik Hekman}.} \bibinfo{year}{2024}\natexlab{}.
\newblock \showarticletitle{The news framing of artificial intelligence: a critical exploration of how media discourses make sense of automation}.
\newblock \bibinfo{journal}{\emph{AI \& SOCIETY}} \bibinfo{volume}{39}, \bibinfo{number}{2} (\bibinfo{date}{April} \bibinfo{year}{2024}), \bibinfo{pages}{437--451}.
\newblock
\showISSN{0951-5666, 1435-5655}
\urldef\tempurl%
\url{https://doi.org/10.1007/s00146-022-01511-1}
\showDOI{\tempurl}


\bibitem[Nisbet et~al\mbox{.}(2002)]%
        {nisbet_knowledge_2002}
\bibfield{author}{\bibinfo{person}{Matthew~C. Nisbet}, \bibinfo{person}{Dietram~A. Scheufele}, \bibinfo{person}{James Shanahan}, \bibinfo{person}{Patricia Moy}, \bibinfo{person}{Dominique Brossard}, {and} \bibinfo{person}{Bruce~V. Lewenstein}.} \bibinfo{year}{2002}\natexlab{}.
\newblock \showarticletitle{Knowledge, {Reservations}, or {Promise}?: {A} {Media} {Effects} {Model} for {Public} {Perceptions} of {Science} and {Technology}}.
\newblock \bibinfo{journal}{\emph{Communication Research}} \bibinfo{volume}{29}, \bibinfo{number}{5} (\bibinfo{date}{Oct.} \bibinfo{year}{2002}), \bibinfo{pages}{584--608}.
\newblock
\showISSN{0093-6502, 1552-3810}
\urldef\tempurl%
\url{https://doi.org/10.1177/009365002236196}
\showDOI{\tempurl}


\bibitem[Observatory(nd)]%
        {oecd_ai_incidents}
\bibfield{author}{\bibinfo{person}{OECD AI~Policy Observatory}.} \bibinfo{year}{n.d.}\natexlab{}.
\newblock \bibinfo{title}{AI Incidents Database}.
\newblock
\newblock
\urldef\tempurl%
\url{https://oecd.ai/en/incidents}
\showURL{%
\tempurl}
\newblock
\shownote{Accessed: 2025-01-17}.


\bibitem[of~the United Nations High Commissioner~for Human Rights~(OHCHR)(2023)]%
        {ohchr2023}
\bibfield{author}{\bibinfo{person}{Office of~the United Nations High Commissioner~for Human Rights~(OHCHR)}.} \bibinfo{year}{2023}\natexlab{}.
\newblock \bibinfo{title}{A Taxonomy of Generative AI and Human Rights Harms}.
\newblock
\newblock
\urldef\tempurl%
\url{https://www.ohchr.org/sites/default/files/documents/issues/business/b-tech/taxonomy-GenAI-Human-Rights-Harms.pdf}
\showURL{%
\tempurl}
\newblock
\shownote{Accessed: 2025-01-17}.


\bibitem[Orwat et~al\mbox{.}(2024)]%
        {orwat_normative_2024}
\bibfield{author}{\bibinfo{person}{Carsten Orwat}, \bibinfo{person}{Jascha Bareis}, \bibinfo{person}{Anja Folberth}, \bibinfo{person}{Jutta Jahnel}, {and} \bibinfo{person}{Christian Wadephul}.} \bibinfo{year}{2024}\natexlab{}.
\newblock \showarticletitle{Normative {Challenges} of {Risk} {Regulation} of {Artificial} {Intelligence}}.
\newblock \bibinfo{journal}{\emph{NanoEthics}} \bibinfo{volume}{18}, \bibinfo{number}{2} (\bibinfo{date}{Aug.} \bibinfo{year}{2024}), \bibinfo{pages}{11}.
\newblock
\showISSN{1871-4757, 1871-4765}
\urldef\tempurl%
\url{https://doi.org/10.1007/s11569-024-00454-9}
\showDOI{\tempurl}


\bibitem[Ou-Yang(2025)]%
        {newspaper2025}
\bibfield{author}{\bibinfo{person}{Lucas Ou-Yang}.} \bibinfo{year}{2025}\natexlab{}.
\newblock \bibinfo{title}{newspaper: Simplified Python Article Scraping \& Curation}.
\newblock
\newblock
\urldef\tempurl%
\url{https://github.com/codelucas/newspaper}
\showURL{%
\tempurl}
\newblock
\shownote{Accessed on January 17, 2025}.


\bibitem[Ouchchy et~al\mbox{.}(2020)]%
        {ouchchy_ai_2020}
\bibfield{author}{\bibinfo{person}{Leila Ouchchy}, \bibinfo{person}{Allen Coin}, {and} \bibinfo{person}{Veljko Dubljević}.} \bibinfo{year}{2020}\natexlab{}.
\newblock \showarticletitle{{AI} in the headlines: the portrayal of the ethical issues of artificial intelligence in the media}.
\newblock \bibinfo{journal}{\emph{AI \& SOCIETY}} \bibinfo{volume}{35}, \bibinfo{number}{4} (\bibinfo{date}{Dec.} \bibinfo{year}{2020}), \bibinfo{pages}{927--936}.
\newblock
\showISSN{0951-5666, 1435-5655}
\urldef\tempurl%
\url{https://doi.org/10.1007/s00146-020-00965-5}
\showDOI{\tempurl}


\bibitem[Paeth et~al\mbox{.}(2024)]%
        {paeth2024lessons}
\bibfield{author}{\bibinfo{person}{Kevin Paeth}, \bibinfo{person}{Daniel Atherton}, \bibinfo{person}{Nikiforos Pittaras}, \bibinfo{person}{Heather Frase}, {and} \bibinfo{person}{Sean McGregor}.} \bibinfo{year}{2024}\natexlab{}.
\newblock \showarticletitle{Lessons for Editors of AI Incidents from the AI Incident Database}.
\newblock \bibinfo{journal}{\emph{arXiv preprint arXiv:2409.16425}} (\bibinfo{year}{2024}).
\newblock


\bibitem[Pang et~al\mbox{.}(2024)]%
        {pang2024blip}
\bibfield{author}{\bibinfo{person}{Rock~Yuren Pang}, \bibinfo{person}{Sebastin Santy}, \bibinfo{person}{Ren{\'e} Just}, {and} \bibinfo{person}{Katharina Reinecke}.} \bibinfo{year}{2024}\natexlab{}.
\newblock \showarticletitle{BLIP: Facilitating the Exploration of Undesirable Consequences of Digital Technologies}. In \bibinfo{booktitle}{\emph{Proceedings of the CHI Conference on Human Factors in Computing Systems}}. \bibinfo{pages}{1--18}.
\newblock


\bibitem[Park et~al\mbox{.}(2024)]%
        {park2024ai}
\bibfield{author}{\bibinfo{person}{Peter~S Park}, \bibinfo{person}{Simon Goldstein}, \bibinfo{person}{Aidan O’Gara}, \bibinfo{person}{Michael Chen}, {and} \bibinfo{person}{Dan Hendrycks}.} \bibinfo{year}{2024}\natexlab{}.
\newblock \showarticletitle{AI deception: A survey of examples, risks, and potential solutions}.
\newblock \bibinfo{journal}{\emph{Patterns}} \bibinfo{volume}{5}, \bibinfo{number}{5} (\bibinfo{year}{2024}).
\newblock


\bibitem[Pasquale(2023)]%
        {pasquale_power_2023}
\bibfield{author}{\bibinfo{person}{Frank Pasquale}.} \bibinfo{year}{2023}\natexlab{}.
\newblock \showarticletitle{Power and {Knowledge} in {Policy} {Evaluation}: {From} {Managing} {Budgets} to {Analyzing} {Scenarios}}.
\newblock \bibinfo{journal}{\emph{Law and Contemporary Problems}} \bibinfo{volume}{86}, \bibinfo{number}{3} (\bibinfo{year}{2023}).
\newblock


\bibitem[Poortvliet et~al\mbox{.}(2016)]%
        {poortvliet_performativity_2016}
\bibfield{author}{\bibinfo{person}{P~Marijn Poortvliet}, \bibinfo{person}{Martijn Duineveld}, {and} \bibinfo{person}{Kai Purnhagen}.} \bibinfo{year}{2016}\natexlab{}.
\newblock \showarticletitle{Performativity in action: {How} risk communication interacts in risk regulation}.
\newblock \bibinfo{journal}{\emph{European Journal of Risk Regulation}} \bibinfo{volume}{7}, \bibinfo{number}{1} (\bibinfo{year}{2016}), \bibinfo{pages}{213--217}.
\newblock
\newblock
\shownote{Publisher: Cambridge University Press}.


\bibitem[Rao et~al\mbox{.}(2024)]%
        {rao2024quallm}
\bibfield{author}{\bibinfo{person}{Varun~Nagaraj Rao}, \bibinfo{person}{Eesha Agarwal}, \bibinfo{person}{Samantha Dalal}, \bibinfo{person}{Dan Calacci}, {and} \bibinfo{person}{Andr{\'e}s Monroy-Hern{\'a}ndez}.} \bibinfo{year}{2024}\natexlab{}.
\newblock \showarticletitle{QuaLLM: An LLM-based Framework to Extract Quantitative Insights from Online Forums}.
\newblock \bibinfo{journal}{\emph{arXiv preprint arXiv:2405.05345}} (\bibinfo{year}{2024}).
\newblock


\bibitem[Reimers and Gurevych(2019)]%
        {reimers-2019-sentence-bert}
\bibfield{author}{\bibinfo{person}{Nils Reimers} {and} \bibinfo{person}{Iryna Gurevych}.} \bibinfo{year}{2019}\natexlab{}.
\newblock \showarticletitle{Sentence-BERT: Sentence Embeddings using Siamese BERT-Networks}. In \bibinfo{booktitle}{\emph{Proceedings of the 2019 Conference on Empirical Methods in Natural Language Processing}}. \bibinfo{publisher}{Association for Computational Linguistics}.
\newblock
\urldef\tempurl%
\url{https://arxiv.org/abs/1908.10084}
\showURL{%
\tempurl}


\bibitem[Reuel et~al\mbox{.}(2024)]%
        {reuel2024open}
\bibfield{author}{\bibinfo{person}{Anka Reuel}, \bibinfo{person}{Ben Bucknall}, \bibinfo{person}{Stephen Casper}, \bibinfo{person}{Tim Fist}, \bibinfo{person}{Lisa Soder}, \bibinfo{person}{Onni Aarne}, \bibinfo{person}{Lewis Hammond}, \bibinfo{person}{Lujain Ibrahim}, \bibinfo{person}{Alan Chan}, \bibinfo{person}{Peter Wills}, {et~al\mbox{.}}} \bibinfo{year}{2024}\natexlab{}.
\newblock \showarticletitle{Open problems in technical ai governance}.
\newblock \bibinfo{journal}{\emph{arXiv preprint arXiv:2407.14981}} (\bibinfo{year}{2024}).
\newblock


\bibitem[Review(2023)]%
        {technologyreview2023}
\bibfield{author}{\bibinfo{person}{MIT~Technology Review}.} \bibinfo{year}{2023}\natexlab{}.
\newblock \bibinfo{title}{The Great Acceleration: CIO Perspectives on Generative AI}.
\newblock
\newblock
\urldef\tempurl%
\url{https://www.technologyreview.com/2023/07/18/1076423/the-great-acceleration-cio-perspectives-on-generative-ai/}
\showURL{%
\tempurl}
\newblock
\shownote{Accessed: 2025-01-17}.


\bibitem[Roberts et~al\mbox{.}(2021)]%
        {roberts2021media}
\bibfield{author}{\bibinfo{person}{Hal Roberts}, \bibinfo{person}{Rahul Bhargava}, \bibinfo{person}{Linas Valiukas}, \bibinfo{person}{Dennis Jen}, \bibinfo{person}{Momin~M Malik}, \bibinfo{person}{Cindy~Sherman Bishop}, \bibinfo{person}{Emily~B Ndulue}, \bibinfo{person}{Aashka Dave}, \bibinfo{person}{Justin Clark}, \bibinfo{person}{Bruce Etling}, {et~al\mbox{.}}} \bibinfo{year}{2021}\natexlab{}.
\newblock \showarticletitle{Media cloud: Massive open source collection of global news on the open web}. In \bibinfo{booktitle}{\emph{Proceedings of the International AAAI Conference on Web and Social Media}}, Vol.~\bibinfo{volume}{15}. \bibinfo{pages}{1034--1045}.
\newblock


\bibitem[Roche et~al\mbox{.}(2022)]%
        {roche_ethics_2022}
\bibfield{author}{\bibinfo{person}{Cathy Roche}, \bibinfo{person}{P.~J. Wall}, {and} \bibinfo{person}{Dave Lewis}.} \bibinfo{year}{2022}\natexlab{}.
\newblock \showarticletitle{Ethics and diversity in artificial intelligence policies, strategies and initiatives}.
\newblock \bibinfo{journal}{\emph{AI and Ethics}} (\bibinfo{date}{Oct.} \bibinfo{year}{2022}).
\newblock
\showISSN{2730-5953, 2730-5961}
\urldef\tempurl%
\url{https://doi.org/10.1007/s43681-022-00218-9}
\showDOI{\tempurl}


\bibitem[Roe and Perkins(2023)]%
        {roe_what_2023}
\bibfield{author}{\bibinfo{person}{Jasper Roe} {and} \bibinfo{person}{Mike Perkins}.} \bibinfo{year}{2023}\natexlab{}.
\newblock \showarticletitle{‘{What} they’re not telling you about {ChatGPT}’: exploring the discourse of {AI} in {UK} news media headlines}.
\newblock \bibinfo{journal}{\emph{Humanities and Social Sciences Communications}} \bibinfo{volume}{10}, \bibinfo{number}{1} (\bibinfo{date}{Oct.} \bibinfo{year}{2023}), \bibinfo{pages}{753}.
\newblock
\showISSN{2662-9992}
\urldef\tempurl%
\url{https://doi.org/10.1057/s41599-023-02282-w}
\showDOI{\tempurl}


\bibitem[Schattschneider(1957)]%
        {schattschneider1957intensity}
\bibfield{author}{\bibinfo{person}{Elmer~E Schattschneider}.} \bibinfo{year}{1957}\natexlab{}.
\newblock \showarticletitle{Intensity, visibility, direction and scope}.
\newblock \bibinfo{journal}{\emph{American Political Science Review}} \bibinfo{volume}{51}, \bibinfo{number}{4} (\bibinfo{year}{1957}), \bibinfo{pages}{933--942}.
\newblock


\bibitem[Scheufele and Lewenstein(2005)]%
        {scheufele_public_2005}
\bibfield{author}{\bibinfo{person}{Dietram~A. Scheufele} {and} \bibinfo{person}{Bruce~V. Lewenstein}.} \bibinfo{year}{2005}\natexlab{}.
\newblock \showarticletitle{The {Public} and {Nanotechnology}: {How} {Citizens} {Make} {Sense} of {Emerging} {Technologies}}.
\newblock \bibinfo{journal}{\emph{Journal of Nanoparticle Research}} \bibinfo{volume}{7}, \bibinfo{number}{6} (\bibinfo{date}{Dec.} \bibinfo{year}{2005}), \bibinfo{pages}{659--667}.
\newblock
\showISSN{1388-0764, 1572-896X}
\urldef\tempurl%
\url{https://doi.org/10.1007/s11051-005-7526-2}
\showDOI{\tempurl}


\bibitem[Schäfer(2017)]%
        {jamieson_how_2017}
\bibfield{author}{\bibinfo{person}{Mike~S. Schäfer}.} \bibinfo{year}{2017}\natexlab{}.
\newblock \bibinfo{booktitle}{\emph{How {Changing} {Media} {Structures} {Are} {Affecting} {Science} {News} {Coverage}}}. Vol.~\bibinfo{volume}{1}.
\newblock \bibinfo{publisher}{Oxford University Press}.
\newblock
\urldef\tempurl%
\url{https://doi.org/10.1093/oxfordhb/9780190497620.013.5}
\showDOI{\tempurl}


\bibitem[Shelby et~al\mbox{.}(2023)]%
        {shelby2023sociotechnical}
\bibfield{author}{\bibinfo{person}{Renee Shelby}, \bibinfo{person}{Shalaleh Rismani}, \bibinfo{person}{Kathryn Henne}, \bibinfo{person}{AJung Moon}, \bibinfo{person}{Negar Rostamzadeh}, \bibinfo{person}{Paul Nicholas}, \bibinfo{person}{N'Mah Yilla-Akbari}, \bibinfo{person}{Jess Gallegos}, \bibinfo{person}{Andrew Smart}, \bibinfo{person}{Emilio Garcia}, {et~al\mbox{.}}} \bibinfo{year}{2023}\natexlab{}.
\newblock \showarticletitle{Sociotechnical harms of algorithmic systems: Scoping a taxonomy for harm reduction}. In \bibinfo{booktitle}{\emph{Proceedings of the 2023 AAAI/ACM Conference on AI, Ethics, and Society}}. \bibinfo{pages}{723--741}.
\newblock


\bibitem[Slattery et~al\mbox{.}(2024)]%
        {slattery2024ai}
\bibfield{author}{\bibinfo{person}{Peter Slattery}, \bibinfo{person}{Alexander~K Saeri}, \bibinfo{person}{Emily~AC Grundy}, \bibinfo{person}{Jess Graham}, \bibinfo{person}{Michael Noetel}, \bibinfo{person}{Risto Uuk}, \bibinfo{person}{James Dao}, \bibinfo{person}{Soroush Pour}, \bibinfo{person}{Stephen Casper}, {and} \bibinfo{person}{Neil Thompson}.} \bibinfo{year}{2024}\natexlab{}.
\newblock \showarticletitle{The ai risk repository: A comprehensive meta-review, database, and taxonomy of risks from artificial intelligence}.
\newblock \bibinfo{journal}{\emph{arXiv preprint arXiv:2408.12622}} (\bibinfo{year}{2024}).
\newblock


\bibitem[Solaiman et~al\mbox{.}(2023)]%
        {solaiman_evaluating_2023}
\bibfield{author}{\bibinfo{person}{Irene Solaiman}, \bibinfo{person}{Zeerak Talat}, \bibinfo{person}{William Agnew}, \bibinfo{person}{Lama Ahmad}, \bibinfo{person}{Dylan Baker}, \bibinfo{person}{Su~Lin Blodgett}, \bibinfo{person}{Hal Daumé~III}, \bibinfo{person}{Jesse Dodge}, \bibinfo{person}{Ellie Evans}, \bibinfo{person}{Sara Hooker}, \bibinfo{person}{Yacine Jernite}, \bibinfo{person}{Alexandra~Sasha Luccioni}, \bibinfo{person}{Alberto Lusoli}, \bibinfo{person}{Margaret Mitchell}, \bibinfo{person}{Jessica Newman}, \bibinfo{person}{Marie-Therese Png}, \bibinfo{person}{Andrew Strait}, {and} \bibinfo{person}{Apostol Vassilev}.} \bibinfo{year}{2023}\natexlab{}.
\newblock \bibinfo{title}{Evaluating the {Social} {Impact} of {Generative} {AI} {Systems} in {Systems} and {Society}}.
\newblock
\newblock
\urldef\tempurl%
\url{http://arxiv.org/abs/2306.05949}
\showURL{%
\tempurl}
\newblock
\shownote{arXiv:2306.05949 [cs]}.


\bibitem[Stahl et~al\mbox{.}(2023)]%
        {stahl_systematic_2023}
\bibfield{author}{\bibinfo{person}{Bernd~Carsten Stahl}, \bibinfo{person}{Josephina Antoniou}, \bibinfo{person}{Nitika Bhalla}, \bibinfo{person}{Laurence Brooks}, \bibinfo{person}{Philip Jansen}, \bibinfo{person}{Blerta Lindqvist}, \bibinfo{person}{Alexey Kirichenko}, \bibinfo{person}{Samuel Marchal}, \bibinfo{person}{Rowena Rodrigues}, \bibinfo{person}{Nicole Santiago}, \bibinfo{person}{Zuzanna Warso}, {and} \bibinfo{person}{David Wright}.} \bibinfo{year}{2023}\natexlab{}.
\newblock \showarticletitle{A systematic review of artificial intelligence impact assessments}.
\newblock \bibinfo{journal}{\emph{Artificial Intelligence Review}} (\bibinfo{date}{March} \bibinfo{year}{2023}).
\newblock
\showISSN{0269-2821, 1573-7462}
\urldef\tempurl%
\url{https://doi.org/10.1007/s10462-023-10420-8}
\showDOI{\tempurl}


\bibitem[Ulken(2005)]%
        {ulken2005question}
\bibfield{author}{\bibinfo{person}{A Ulken}.} \bibinfo{year}{2005}\natexlab{}.
\newblock \bibinfo{title}{Question of Balance: Are Google News search results politically biased}.
\newblock
\newblock


\bibitem[Uuk et~al\mbox{.}({[n.\,d.]})]%
        {uuk_taxonomy_nodate}
\bibfield{author}{\bibinfo{person}{Risto Uuk}, \bibinfo{person}{Carlos~Ignacio Gutierrez}, \bibinfo{person}{Daniel Guppy}, \bibinfo{person}{Lode Lauwaert}, \bibinfo{person}{Lucia Velasco}, \bibinfo{person}{Peter Slattery}, {and} \bibinfo{person}{Carina Prunkl}.} \bibinfo{year}{[n.\,d.]}\natexlab{}.
\newblock \showarticletitle{A {Taxonomy} of {Systemic} {Risks} from {General}-{Purpose} {AI}}.
\newblock  (\bibinfo{year}{[n.\,d.]}).
\newblock


\bibitem[Vergeer(2020)]%
        {vergeer_artificial_2020}
\bibfield{author}{\bibinfo{person}{Maurice Vergeer}.} \bibinfo{year}{2020}\natexlab{}.
\newblock \showarticletitle{Artificial {Intelligence} in the {Dutch} {Press}: {An} {Analysis} of {Topics} and {Trends}}.
\newblock \bibinfo{journal}{\emph{Communication Studies}} \bibinfo{volume}{71}, \bibinfo{number}{3} (\bibinfo{date}{May} \bibinfo{year}{2020}), \bibinfo{pages}{373--392}.
\newblock
\showISSN{1051-0974, 1745-1035}
\urldef\tempurl%
\url{https://doi.org/10.1080/10510974.2020.1733038}
\showDOI{\tempurl}


\bibitem[Wakunuma and Eke(2024)]%
        {wakunuma_africa_2024}
\bibfield{author}{\bibinfo{person}{Kutoma Wakunuma} {and} \bibinfo{person}{Damian Eke}.} \bibinfo{year}{2024}\natexlab{}.
\newblock \showarticletitle{Africa, {ChatGPT}, and {Generative} {AI} {Systems}: {Ethical} {Benefits}, {Concerns}, and the {Need} for {Governance}}.
\newblock \bibinfo{journal}{\emph{Philosophies}} \bibinfo{volume}{9}, \bibinfo{number}{3} (\bibinfo{date}{June} \bibinfo{year}{2024}), \bibinfo{pages}{80}.
\newblock
\showISSN{2409-9287}
\urldef\tempurl%
\url{https://doi.org/10.3390/philosophies9030080}
\showDOI{\tempurl}


\bibitem[Ward et~al\mbox{.}(2013)]%
        {ward2013comparing}
\bibfield{author}{\bibinfo{person}{Michael~D Ward}, \bibinfo{person}{Andreas Beger}, \bibinfo{person}{Josh Cutler}, \bibinfo{person}{Matthew Dickenson}, \bibinfo{person}{Cassy Dorff}, {and} \bibinfo{person}{Ben Radford}.} \bibinfo{year}{2013}\natexlab{}.
\newblock \showarticletitle{Comparing GDELT and ICEWS event data}.
\newblock \bibinfo{journal}{\emph{Analysis}} \bibinfo{volume}{21}, \bibinfo{number}{1} (\bibinfo{year}{2013}), \bibinfo{pages}{267--297}.
\newblock


\bibitem[Weidinger et~al\mbox{.}(2023)]%
        {weidinger2023sociotechnical}
\bibfield{author}{\bibinfo{person}{Laura Weidinger}, \bibinfo{person}{Maribeth Rauh}, \bibinfo{person}{Nahema Marchal}, \bibinfo{person}{Arianna Manzini}, \bibinfo{person}{Lisa~Anne Hendricks}, \bibinfo{person}{Juan Mateos-Garcia}, \bibinfo{person}{Stevie Bergman}, \bibinfo{person}{Jackie Kay}, \bibinfo{person}{Conor Griffin}, \bibinfo{person}{Ben Bariach}, {et~al\mbox{.}}} \bibinfo{year}{2023}\natexlab{}.
\newblock \showarticletitle{Sociotechnical safety evaluation of generative ai systems}.
\newblock \bibinfo{journal}{\emph{arXiv preprint arXiv:2310.11986}} (\bibinfo{year}{2023}).
\newblock


\bibitem[Weidinger et~al\mbox{.}({[n.\,d.]})]%
        {weidinger_taxonomy_nodate}
\bibfield{author}{\bibinfo{person}{Laura Weidinger}, \bibinfo{person}{Jonathan Uesato}, \bibinfo{person}{Maribeth Rauh}, \bibinfo{person}{Conor Griffin}, \bibinfo{person}{Po-Sen Huang}, \bibinfo{person}{John Mellor}, \bibinfo{person}{Amelia Glaese}, \bibinfo{person}{Myra Cheng}, \bibinfo{person}{Borja Balle}, \bibinfo{person}{Atoosa Kasirzadeh}, \bibinfo{person}{Courtney Biles}, \bibinfo{person}{Sasha Brown}, \bibinfo{person}{Zac Kenton}, \bibinfo{person}{Will Hawkins}, \bibinfo{person}{Tom Stepleton}, \bibinfo{person}{Abeba Birhane}, \bibinfo{person}{Lisa~Anne Hendricks}, \bibinfo{person}{Laura Rimell}, \bibinfo{person}{William Isaac}, \bibinfo{person}{Julia Haas}, \bibinfo{person}{Sean Legassick}, \bibinfo{person}{Geoffrey Irving}, {and} \bibinfo{person}{Iason Gabriel}.} \bibinfo{year}{[n.\,d.]}\natexlab{}.
\newblock \bibinfo{title}{Taxonomy of {Risks} posed by {Language} {Models}}.
\newblock
\newblock
\urldef\tempurl%
\url{https://doi.org/10.1145/3531146.3533088}
\showDOI{\tempurl}
\newblock
\shownote{Publication Title: 2022 ACM Conference on Fairness, Accountability, and Transparency Volume: 22}.


\bibitem[Yang and Menczer(2023)]%
        {yang2023anatomy}
\bibfield{author}{\bibinfo{person}{Kai-Cheng Yang} {and} \bibinfo{person}{Filippo Menczer}.} \bibinfo{year}{2023}\natexlab{}.
\newblock \showarticletitle{Anatomy of an AI-powered malicious social botnet}.
\newblock \bibinfo{journal}{\emph{arXiv preprint arXiv:2307.16336}} (\bibinfo{year}{2023}).
\newblock


\bibitem[Zeng et~al\mbox{.}(2024)]%
        {zeng_ai_2024}
\bibfield{author}{\bibinfo{person}{Yi Zeng}, \bibinfo{person}{Kevin Klyman}, \bibinfo{person}{Andy Zhou}, \bibinfo{person}{Yu Yang}, \bibinfo{person}{Minzhou Pan}, \bibinfo{person}{Ruoxi Jia}, \bibinfo{person}{Dawn Song}, \bibinfo{person}{Percy Liang}, {and} \bibinfo{person}{Bo Li}.} \bibinfo{year}{2024}\natexlab{}.
\newblock \bibinfo{title}{{AI} {Risk} {Categorization} {Decoded} ({AIR} 2024): {From} {Government} {Regulations} to {Corporate} {Policies}}.
\newblock
\newblock
\urldef\tempurl%
\url{https://doi.org/10.48550/ARXIV.2406.17864}
\showDOI{\tempurl}
\newblock
\shownote{Version Number: 1}.


\bibitem[Zhang et~al\mbox{.}(2023)]%
        {zhang2023safetybench}
\bibfield{author}{\bibinfo{person}{Zhexin Zhang}, \bibinfo{person}{Leqi Lei}, \bibinfo{person}{Lindong Wu}, \bibinfo{person}{Rui Sun}, \bibinfo{person}{Yongkang Huang}, \bibinfo{person}{Chong Long}, \bibinfo{person}{Xiao Liu}, \bibinfo{person}{Xuanyu Lei}, \bibinfo{person}{Jie Tang}, {and} \bibinfo{person}{Minlie Huang}.} \bibinfo{year}{2023}\natexlab{}.
\newblock \showarticletitle{Safetybench: Evaluating the safety of large language models with multiple choice questions}.
\newblock \bibinfo{journal}{\emph{arXiv preprint arXiv:2309.07045}} (\bibinfo{year}{2023}).
\newblock


\end{thebibliography}
